\documentclass{aa}

\usepackage{graphicx}
\usepackage{txfonts}
\usepackage{float}
\usepackage{mathtools}
\usepackage{subcaption}
\usepackage[version=4]{mhchem}
\usepackage{gensymb}
\usepackage{tikzsymbols}
\usepackage{rotating}
\usepackage{multirow}
\usepackage{makecell}
\usepackage{mleftright}
\usepackage{ulem}

     \def\aap{A\&A}
     \def\apjl{ApJL}
     \def\apjs{ApJS}

\begin{document} 

   \title{Clouds form on the hot Saturn  JWST ERO target WASP-96b}
   \author{
         D. Samra\inst{1}
         \and
   Ch. Helling \inst{1,2}
   \and
      K. L. Chubb\inst{3}
   \and
   M. Min\inst{4}
   \and
    L. Carone\inst{1}
    \and
    A. D. Schneider\inst{5,6}
     }
   
 \institute{
      Space Research Institute, Austrian Academy of Sciences, Schmiedlstrasse 6, A-8042 Graz, Austria\\
     \email{Dominic.Samra@oeaw.ac.at}    
            \and
            Institute for Theoretical Physics and Computational Physics, Graz University of Technology, Petersgasse 16
8010 Graz
\and
Center for Exoplanet Science, University of St Andrews, KY16 9AJ, UK
         \and
         SRON Netherlands Institute for Space Research, Niels Bohrweg 4, 2333 CA Leiden, Netherlands
             \and
             Institute of Astronomy, KU Leuven, Celestijnenlaan 200D, 3001, Leuven, Belgium
            \and
             Center for ExoLife Sciences, Niels Bohr Institute, Øster Voldgade 5, 1350 Copenhagen, Denmark
             }

      \date{\today, Received September 15, 2996; accepted March 16, 2997}

\abstract
    {WASP-96b is a hot Saturn exoplanet, with an equilibrium temperature of $\approx\,1300\,{\rm K}$, well within the regime of thermodynamically expected extensive cloud formation. Prior observations with Hubble/WFC3, Spitzer/IRAC, and VLT/FORS2 have been combined into a single spectra for which retrievals suggest  a cold but cloud-free atmosphere. Recently, the planet was observed with the James Webb Space Telescope (JWST) as part of the Early Release Observations (ERO).
    }
    {The formation of clouds in the atmosphere of the exoplanet WASP-96b is explored.}
      {1D profiles are extracted from the 3D GCM expeRT/MITgcm results and used 
     as input for a kinetic, non-equilibrium model to study the formation of mineral cloud particles of mixed composition.  The ARCiS retrieval framework is applied to the pre-JWST WASP-96b transit spectra to investigate the apparent contradiction between cloudy models and assumed cloud-free transit spectra.}
    {Clouds are predicted to be ubiquitous throughout the atmosphere of WASP-96b. Silicate materials contribute between 40\% and 90\%, hence, also metal oxides do contribute with up to 40\% in the low-pressure regimes that effect the spectra. We explore how to match these cloudy models with currently available atmospheric transit spectra. A reduced vertical mixing acts to settle clouds to deeper in the atmosphere, and an increased cloud particles  porosity reduces the opacity of clouds in the near-IR and optical region. Both these effects allow for clearer molecular features to be observed, while still allowing clouds to be in the atmosphere.
    }
    {The atmosphere of WASP-96b is unlikely to be cloud free. Also retrievals of HST, Spitzer and VLT spectra show that multiple cloudy solutions reproduce the data. JWST observations will be affected by clouds, where within even the NIRISS wavelength range the cloud top pressure varies by an order of magnitude. The long wavelength end of NIRSpec and short end of MIRI may probe atmospheric asymmetries between the limbs of the terminator on WASP-96b.}   


  \maketitle

\section{Introduction}

The hot Saturn WASP-96b was one of the first\footnote{HAT-P-14b was observed during commissioning, strictly being the first exoplanet observations released in \cite{Rigby2022_JWST_Commissioning}.} extrasolar planets for which the James Webb Space Telescope (JWST) observations were released to the community on the 12$^{\rm th}$ July 2022. The planet's primary transit was observed using the telescope's NIRISS/SOSS instrument as part of the JWST Early Release Observations (ERO)~\citep{22PoBlBr_arxiv}. It therefore presents itself as a desirable target to analyse from a modelling perspective, in preparation for newly reduced transit spectra expected to become available in the near future. WASP-96b has a mass of 0.48~$\pm$~0.03~M$_{\rm Jup}$, a mean radius of 1.2~$\pm$~0.06~R$_{\rm Jup}$, and orbits a G-type stars in 3.4 days at a distance of 1120 lyr from Earth~\citep{14HeAnCa}. The orbit appears non-eccentric and its global temperature is T$_{\rm glob}\approx 1300$K (1285~$\pm$~40, \citealt{14HeAnCa}). Optical transmission spectra of WASP-96b have been observed pre-JWST by the FORS2 spectrograph on the ground-based Very Large Telescope (VLT)~\citep{2018Natur.557..526N}. The resolution of these VLT spectra allowed the Na doublet to be studied in detail;
Na abundances were derived and a cloud-free atmosphere was assumed because of broad Na I line wings by \citep{2018Natur.557..526N}. \cite{2022MNRAS.tmp.1485N} more recently used space-based instruments to observe further primary transits of WASP-96b, including the Infrared Array Camera (IRAC) on the Spitzer Space Telescope (Spitzer), and the Wide Field Camera 3 (WFC3) instrument on the Hubble Space Telescope (HST). These were combined with ground-based observations from the VLT to do a more complete analysis of the planetary atmosphere. These are the set of observations that we use for some comparisons to our atmospheric models in this work.
\cite{2022MNRAS.tmp.1485N} argue that their retrieval approach of these combined data suggests cloud/haze free terminators at the pressures probed by the observations. The retrieval procedure treats cloud particles rain out and the element abundances for Na, K, C and O as independent parameters. The authors derive solar-to-super-solar oxygen and sodium element abundances. A super-solar oxygen abundance may be the reason for the strong H$_2$O features which the authors present as an additional reason to conclude a cloud-free atmosphere of WASP-96b.
Metallicity derived for the host star WASP-96 are either solar or super-solar; \cite{14HeAnCa} find abundances compared to solar of $0.14\pm0.19$~dex, but more recent observations by \cite{2022MNRAS.tmp.1485N} suggest that WASP-96 has enhanced metallicity.

A recently analysed observation of brown dwarf VHS~1256-1257~b by \cite{22MiBiPa} using JWST's NIRSpec IFU and MIRI MRS modes, revealed solid-states silicate features in the longer wavelength part of the atmospheric spectrum, giving strong evidence for small silicate particles in the atmosphere.

We demonstrate in this paper that mineral cloud particles that are made of a mix of metal-oxide and silicate components are expected to form in the atmosphere of the hot Saturn WASP-96b.
We present a cloud map for this planet which demonstrate its complete coverage with clouds. We identify WASP-96b as sitting right at the boundary between the weather/climate regime for cool planets and the less homogeneous transition planets that were introduced in \cite{Helling2022}.. The absence of a strong morning/evening terminator asymmetry should improve the chances of observing spectral features of the ensemble of cloud particles that our microphysical models predict should populate the terminator regions.

\begin{figure}
\hspace*{-0.7cm}
    \includegraphics[width=1.1\linewidth]{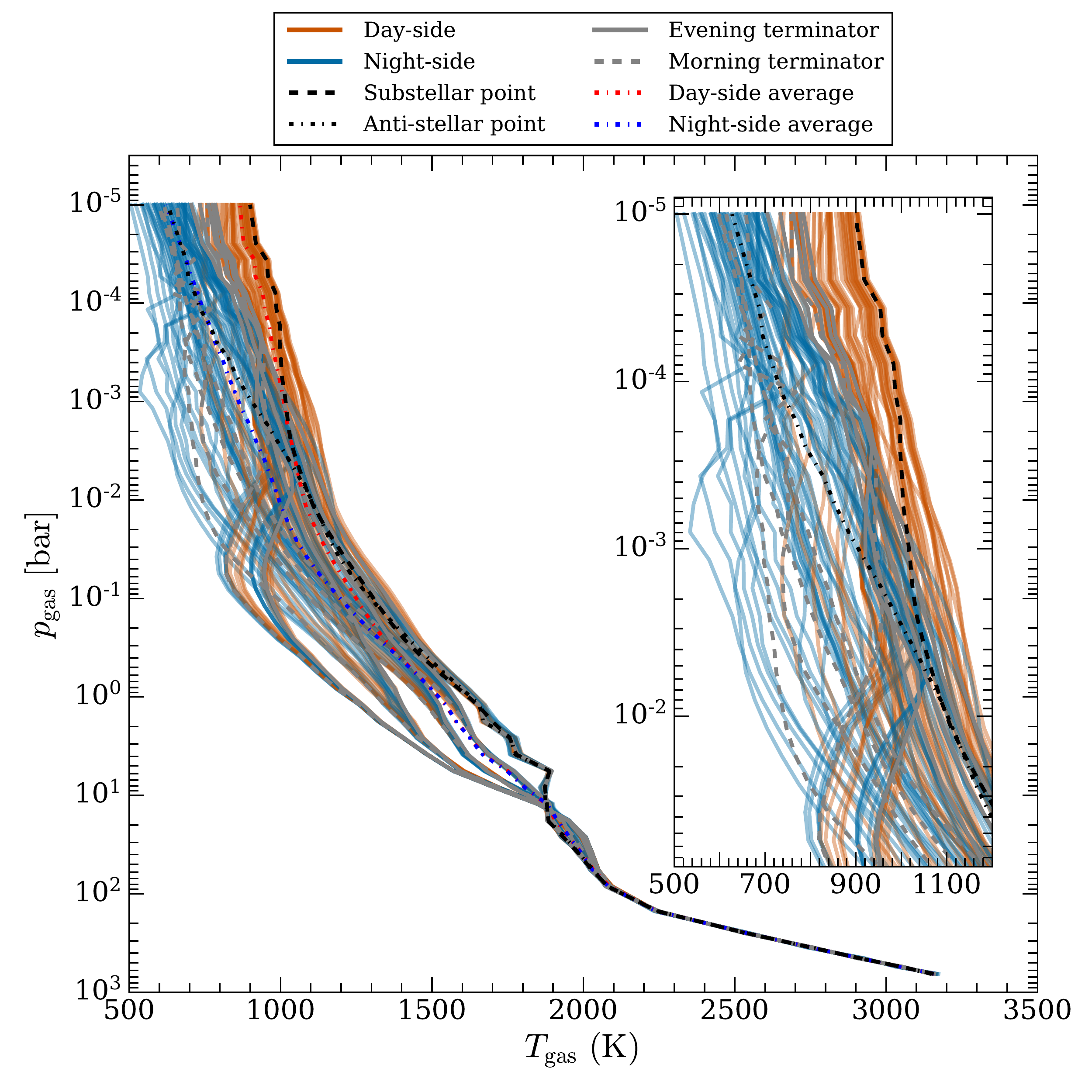}
    \caption{Pressure-temperature structure of the 120 1D profiles extracted from the GCM results. Orange lines represent the dayside profiles, blue lines the nightside profiles, and the grey lines the terminator profiles. Black dashed and dot-dashed lines highlight the anti-stellar points respectively. Day- and night-side hemisphere averages shown in red and blue dashed lines. Inset shows the upper atmosphere GCM of the profiles.}
    \label{fig:1dprofiles}
\end{figure}

\section{Approach}\label{s:ap}
We examine the cloud structure on the hot Saturn WASP-96b by adopting a  hierarchical approach similar to works on the hot Jupiters HD\,189733b and HD\,209458b (\citealt{Lee2015,2016MNRAS.460..855H}), and the ultra-hot Jupiters WASP-18b (\citealt{2019arXiv190108640H}) and HAT-P-7b (\citealt{2019arXiv190608127H,2020A&A...635A..31M}): The first modelling step produces a cloud-free 3D GCM result for a model representing WASP-96b. 
These results are used as input for the second modelling step which is a kinetic cloud formation model consistently combined with equilibrium gas-chemistry calculations. We utilise 120 1D ($T_{\rm gas}$(z), $p_{\rm gas}$(z), $v_{\rm z}(z)$)-profiles for WASP-96b similar to our previous works. $T_{\rm gas}$(z) is the local gas temperature [K], $p_{\rm gas}$(z) is the local gas pressure [bar], and $v_{\rm z}$(z) is the local vertical velocity component [cm s$^{-1}$]. 

Such a hierarchical approach has the limitation of not explicitly taking into account the potential effect of horizontal winds on cloud formation. However, processes governing the formation of mineral clouds are determined by local thermodynamic properties which are the result of 3D dynamic atmosphere  simulations.  Cloud particle properties such as particle size or particle composition could be somewhat smeared out in longitude compared to the results shown here. The temperature structure may change if the cloud particle opacity is fully taken into account in the solution of the radiative transfer. This may change the precise location of the cloud in pressure space but not the principle result of clouds forming in WASP-96b.

\paragraph{3D atmosphere modelling:}
We utilise expeRT/MITgcm \citep{2019arXiv190413334C, 2021MNRAS.tmp.1277B} to model WASP-96b. expeRT/MITgcm builds on the dynamical core of MITgcm and has been adapted to model tidally locked gas giants. Recent extensions in \cite{2022arXiv220209183S} include non-grey radiative transfer coupling. The model parameters used for the GCM, representative of the hot Saturn WASP-96b, are: $R_{\rm p} = 8.58\times10^{9}\,{\rm cm}$, $P_{\rm rot} = 3.4\,{\rm days}$, $\log_{10}(g\,{\rm [cm\,s^{-2}]}) = 2.9$, and the substellar point irradiation temperature $T_{\rm irr} = 1819\,{\rm K}$ (Eq. 20 \citealt{2022arXiv220209183S}). The model is run for $1400\,{\rm days}$, and we note that the deepest layers may not be fully converged. A constant mean molecular weight of $\mu=2.3$ is assumed for the 3D atmospheric modelling. This constant value is a reasonable assumption given that the thermodynamic structure of the atmosphere does not cause the gas composition to deviate from a H$_2$-dominated gas. Additional details for the 3D GCM setup can be found in Table~\ref{tab:GCM_params}.

\cite{2022MNRAS.tmp.1485N} fit an isothermal black-body to the dayside of WASP-96b using {\it Spitzer} observations, finding a dayside brightness temperature of $1545\pm90\,{\rm K}$. The equilibrium temperature of the planet can be calculated using $T_{\rm eq} = \sqrt{R_{\star}/a_{\rm p}}T_{\star}(f(1~-~A_{\rm B}))^{1/4}$, the final factor includes the heat re-distribution factor ($f$) and the bond Albedo of the planet ($A_{\rm B}$). Assuming full heat redistribution ($f=1/4$) and full absorption of incident radiation ($A_{\rm B} = 0$), one arrives at the equilibrium temperature of the planet: $T_{\rm eq} = 1286\,{\rm K}$. As input to the initialisation of the GCM, the substellar irradiation temperature is used, this is the temperature of a black-body re-radiating the energy received from the star, by the planet at the substellar point ($T_{\rm irr} =1819\,{\rm K}$, see Table~\ref{tab:GCM_params}). Comparing these numbers to the observed brightness temperature is difficult, the brightness temperature sits between the planet equilibrium temperature and the substellar irradiation temperature. This indicates some level of heat re-distribution around the planet, as well as the potential impact of cloud albedo. However, observations of similar temperature exoplanets have revealed low dayside albedos (e.g. \cite{Schwartz2015MNRAS.449.4192S,Fraine2021,Brandeker2022}).
  
\begin{figure*}
\includegraphics[page=1,width=0.55\linewidth]{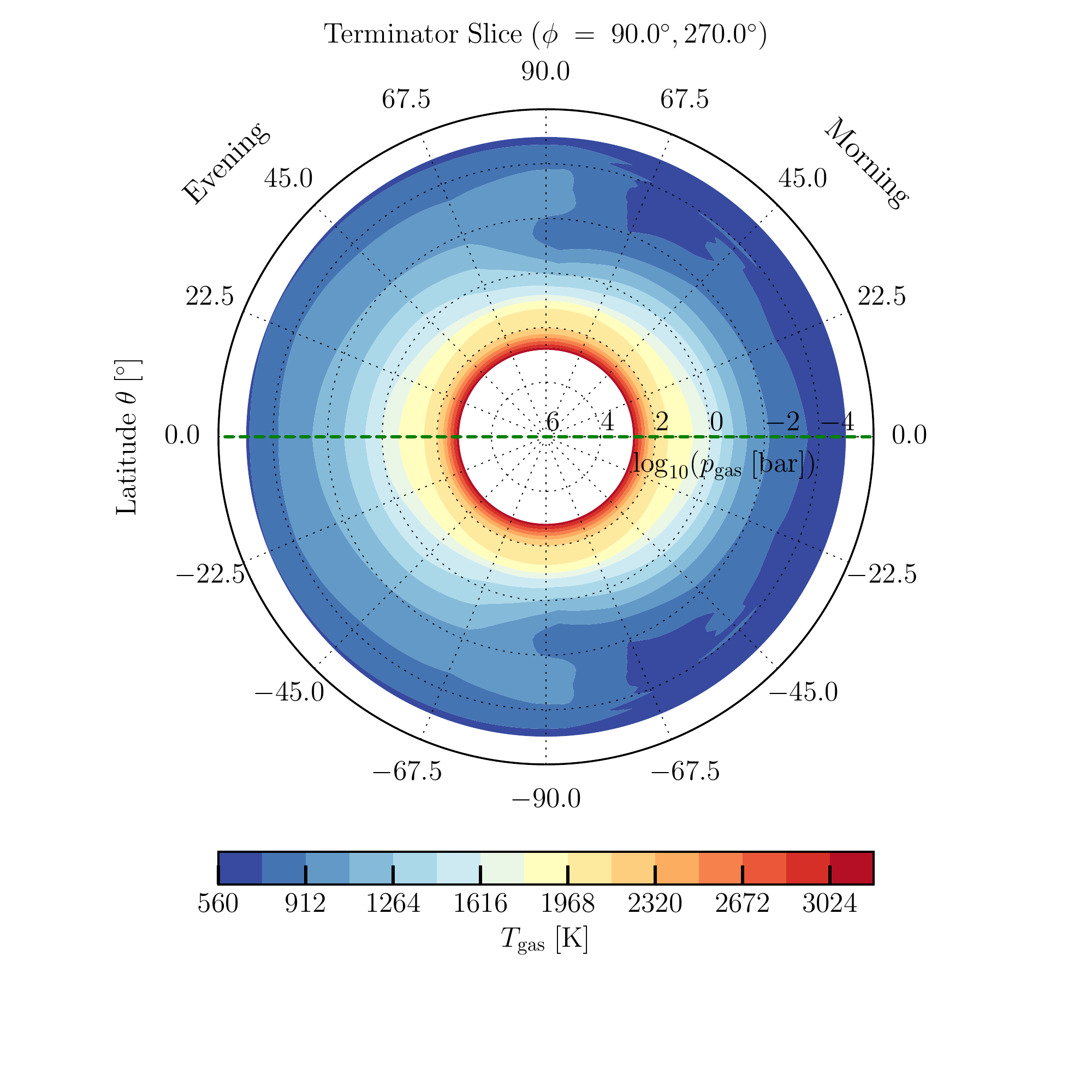}
\hspace*{-1.5cm}\includegraphics[page=4,width=0.55\linewidth]{Figures/Slice_Plots_WASP096b_termonly.pdf}\\*[-1cm]
\includegraphics[page=2,width=0.55\linewidth]{Figures/Slice_Plots_WASP096b_termonly.pdf}
\hspace*{-1.5cm}\includegraphics[page=3,width=0.55\linewidth]{Figures/Slice_Plots_WASP096b_termonly.pdf}
\caption{WASP-96b cloud maps in 2D terminator slices as a result of the microphysical cloud model:
 {\bf Top Left:} Local atmospheric gas temperature and  gas pressure (T$_{\rm gas}$, p$_{\rm gas}$), {\bf Top Right:} Total mineral seed formation (nucleation) rate, $J_*=\sum_i J_{\rm i}$ [cm$^{-3}$ s$^{-1}$] (i=TiO$_2$, SiO, NaCl, KCl). {\bf Bottom left:} Dust to gas ratio $\rho_{\rm d}/ \rho$. {\bf Bottom right:} Mean cloud particle radius $\langle a \rangle\,[\mu {\rm m}]$.}
  \label{TpNuc}
\end{figure*}

\begin{figure*}
\includegraphics[page=1,width=0.55\linewidth]{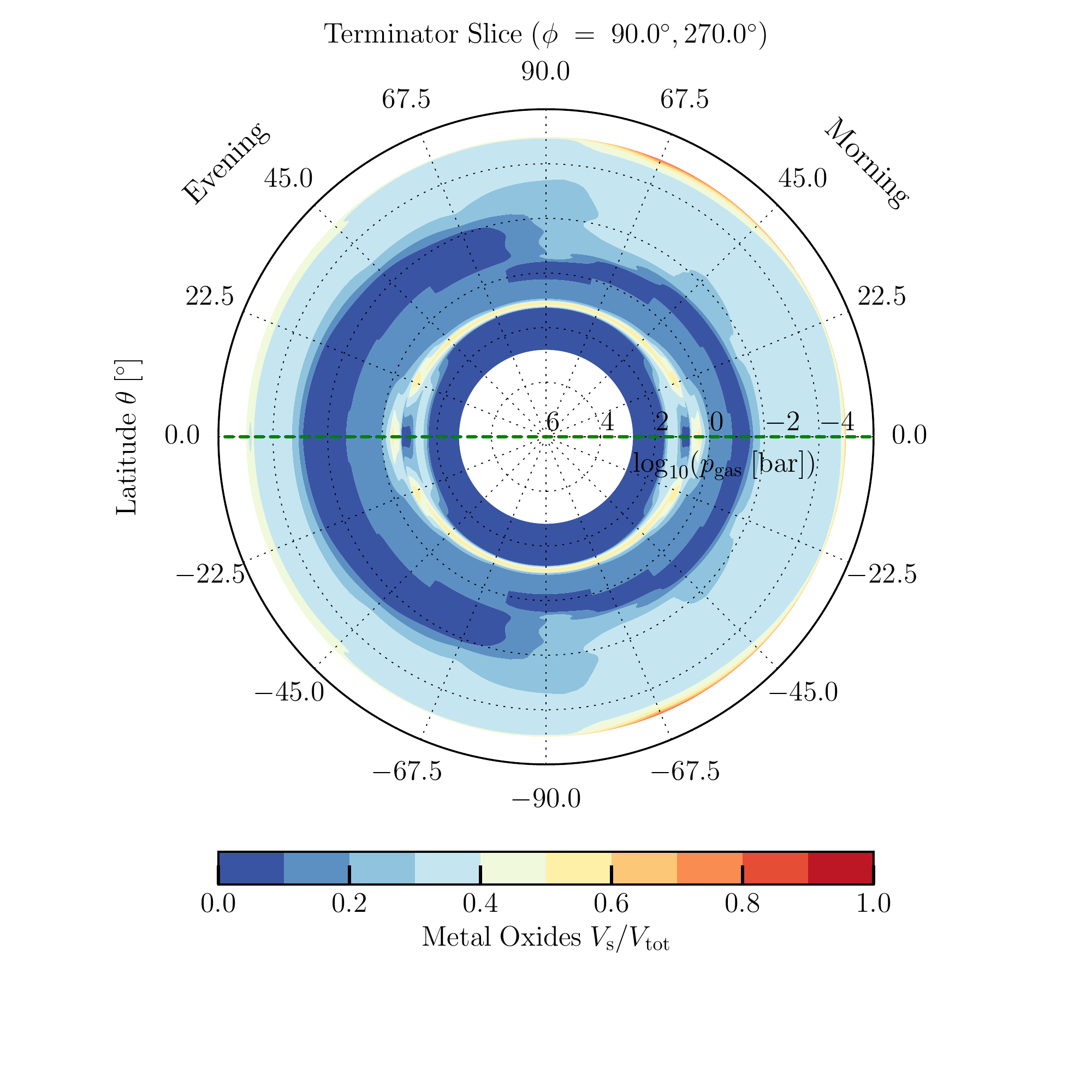}
\hspace*{-1.5cm}\includegraphics[page=2,width=0.55\linewidth]{Figures/Slice_Plots_MaterialComps_WASP096b_termonly.pdf}\\*[-1cm]
\includegraphics[page=3,width=0.55\linewidth]{Figures/Slice_Plots_MaterialComps_WASP096b_termonly.pdf}
\hspace*{-1.5cm}\includegraphics[page=4,width=0.55\linewidth]{Figures/Slice_Plots_MaterialComps_WASP096b_termonly.pdf}
\caption{Volume ratio of the cloud forming materials for WASP-96b in 2D terminator slices, for the same models as in Fig.~\ref{TpNuc}, materials are grouped as in \citep{Helling2021}: {\bf Top Left:} Metal oxides ($s=$\ce{SiO}[s], \ce{SiO2}[s], \ce{MgO}[s], \ce{FeO}[s], \ce{Fe2O3}[s]), {\bf Top Right:} Silicates ($s=$\ce{MgSiO3}[s], \ce{Mg2SiO4}[s], \ce{Fe2SiO4}[s], \ce{CaSiO3}[s]). 
{\bf Bottom left} High temperature condensates ($s=$\ce{TiO2}[s], \ce{Fe}[s], \ce{FeS}[s], \ce{Al2O3}[s], \ce{CaTiO3}[s]), {\bf Bottom right:} Salts ($s=$\ce{KCl}[s], \ce{NaCl}[s].)
}
  \label{VsVtot}
\end{figure*}

\paragraph{Kinetic cloud formation:} We apply the same set-up of our kinetic cloud formation model (nucleation, growth, evaporation, gravitational settling, element consumption and replenishment)  and equilibrium gas-phase calculations as for our grid study in  \cite{Helling2022}. It is essential that the seed forming species are also considered as surface growth material, since both processes (nucleation and growth) compete for the participating elements (Ti, Si, O, Na, K, and Cl in this work). We consider the formation of 16 bulk materials (s=\ce{TiO2}[s], \ce{Mg2SiO4}[s], \ce{MgSiO3}[s], MgO[s], SiO[s], \ce{SiO2}[s], Fe[s], FeO[s], FeS[s], \ce{Fe2O3}[s], \ce{Fe2SiO4}[s], \ce{Al2O3}[s], \ce{CaTiO3}[s], \ce{CaSiO3}[s], KCl[s], NaCl[s]) that form from 11 elements (Mg, Si, Ti, O, Fe, Al, Ca, S, Na, K and Cl) by 132 surface reactions. In total, we are solving 31 ODEs to describe the formation of cloud condensation nuclei ($J_*(z) = \sum_i J_{\rm i}$, i=TiO$_2$, SiO, KCl, NaCl) that grow to macroscopic sized cloud particles made  of a mix of materials which changes depending on the local atmospheric gas temperature and gas pressure.

\paragraph{Retrieval and atmospheric spectra modelling with ARCiS:}
We follow the method of \cite{20MiOrCh.arcis} by running a constrained retrieval on the previously published VLT/HST/Spitzer transmission spectra of WASP-96b (collated by \cite{2022MNRAS.tmp.1485N}) using Bayesian retrieval code ARCiS (ARtful modelling Code for exoplanet Science). The constrained retrieval includes the cloud formation models of \cite{18OrMi.arcis}, with the variety of cloud species included in this work the same as in \cite{20MiOrCh.arcis} with additional Na and K silicate clouds . It is a simplified approach in comparison to the previous hierarchical modelling stages of this work, but it enables us to compare some retrieved parameters to our kinetic cloud formation models, such as cloud particle size, and expected cloud optical thickness.

It also allows us to explore the effect on the spectra of some of these parameters, which we do using the forward modelling mode of ARCiS. We generate simulated transmission spectra with varied atmospheric parameters, which can be compared to observations. Here, we focus on the effects of clouds in the atmosphere and how they can be present but still let us observe strong molecular features. The distribution of hollow spheres (DHS) method is employed in ARCiS for computing the cloud opacities~\citep{05MiHoKo}. This allows for the inclusion of non-spherical cloud particles, and the ability to vary the porosity of the particles. A similar simple model for particle porosity incorporating also the impacts of increased surface area and reduced settling on the clouds as well as optical effects was recently explored in \cite{20SaHeMi}. We also explore the impact of this microphysically consistent model on the cloud optical depth in Section~\ref{subsec:Cloud_opt_depth}. The chemistry for all retrievals is constrained, with elemental abundances computed after the cloud formation process in order to take into account elements which may have formed into clouds. The equilibrium chemistry code GGchem~\citep{2018A&A...614A...1W} is used to compute the molecular abundances given these elemental abundances. A summary of the retrieved parameters and their priors is given in Table~\ref{t:ret_pars}. For comparison, we also run a cloud-free retrieval, with all other parameters the same. We present the results of these retrievals in Section~\ref{subsec:retrieval_results}.

\section{Cloud properties on the hot Saturn WASP-96b}\label{s:clouds_emerging}

\subsection{The 3D GCM atmosphere structure}\label{s:Tp}

The 3D GCM results for the hot Saturn WASP-96b suggest a maximum day/night temperature contrast of $\approx 400$K occurring in the low-pressure regimes at $p_{\rm gas}<10^{-2}$~bar. Figure~\ref{fig:1dprofiles} demonstrates this by the 120 1D profiles extracted from the 3D GCM solution. The morning terminator (grey dashed line) shows some moderate temperature inversion of $\approx 100$~K resulting from the cold gas being transported from the nightside to the dayside. The temperature differences between the morning and the evening terminators are maximum $\approx 200$~K.

Based on its global temperature $T_{\rm glob}\approx 1300$~K, the hot Saturn WASP-96b falls onto the boundary of the cool planet climate regimes identified in \cite{Helling2022}. The cool planet climate regimes are characterised by a globally homogeneous formation of cloud condensation nuclei that leads to a globally homogeneous cloud coverage.  WASP-96b does have some day/night temperature difference as shown in Fig.~\ref{fig:1dprofiles}. The day/night temperature difference does cause a temperature difference in the evening and morning terminator regions as shown in the 2D terminator slice plot  for local gas temperature in Fig.~\ref{TpNuc} (top left). The evening terminator is affected by the warm gas being advected from the dayside to the nightside, the morning terminator by the cold gas coming from the nightside. This causes a slightly elliptically shaped high-temperature area in the equatorial regions ($\theta = 0^{o}$).

\subsection{The mineral clouds on WASP-96b}\label{s:clouds_mineral}

In this section, we discuss the results for mineral cloud formation in the hot Saturn WASP-96b to enable a detailed insight into the cloud properties that are likely to shape the JWST spectrum of the planet.

\subsubsection{Very mildly asymmetric cloud coverage and mildly non-homogeneous cloud properties}

{\it Firstly,} we observe that cloud condensation nuclei (CCNs) form for all the 1D profiles used here to probe cloud formation in the 3D atmosphere of WASP-96b (Fig.~\ref{TpNuc}, top right). This is in stark contrast to the ultra-hot Jupiters, like WASP-18b and HAT-P-7b, but in line with other hot Jupiters like HD\,189733b and HD\,209458\,b. {\it Secondly,} the rate of seed formation differs between the day and the nightside because of the different thermodynamic conditions. The morning terminator regions have a somewhat larger atmospheric volume where efficient formation of CCNs takes place compared to the evening terminator (i.e. at the morning terminator the region of efficient nucleation extends to deeper pressures than at the evening terminator). The nucleation rate, $J_*$, follows the asymmetric morning/evening terminator temperature differences.

Once the CCNs are formed, they grow to macroscopic cloud particles through condensation, these cloud particles fall into the atmosphere towards higher pressures where growth may accelerate, before the cloud particles evaporate where their materials become thermally unstable at the cloud base (at $p_{\rm gas}\approx 10^{2.2}$ bar, consistently around the terminator). Therefore, the extension of the cloud is determined by the atmospheric volume that is populated by cloud particles (Fig.~\ref{TpNuc}, bottom left), not by where the CCNs form (Fig.~\ref{TpNuc}, top right). Mean cloud particle sizes range from 0.3 $\mu$m at $p_{\rm gas}=10^{-2}$bar, to 15$\mu$m at 1bar and 0.1\,mm at the cloud base at $p_{\rm gas}=158$ bar. 

The cloud extension is affected by the vertical mixing efficiency but also by material properties like the cloud particle porosity.  The vertical mixing counteracts the element depletion of the gas phase by cloud formation which would inhibit the continuous presence of clouds in an atmosphere (see Sect. 3.3. in \citealt{Woitke2003}). \cite{20SaHeMi} demonstrated that highly porous (low density) cloud would expand the cloud layer upwards, due to the increased surface area provided by porous cloud particles. However, the cloud base is unaffected as it is determined by thermal stability. High-porosity cloud particle would also have a larger mean particles size and would result in a somewhat larger dust-to-gas ratio.

Figure~\ref{TpNuc} (lower left) shows that the cloud particle mass load (in terms of the dust-to-gas ration $\rho_{\rm d}/\rho$) of the WASP-96b model atmosphere appears rather symmetrically distributed as a result of the homogeneous thermodynamic structure that is suggested by the 3D GCM for WASP-96b. However, there is still a slight asymmetry between the two terminators in the peak cloud particle mass load, due to the increased particle surface area from the larger nucleation rate on the morning terminator.

\subsubsection{General material composition of cloud particles}

\begin{figure}[h!]
    \includegraphics[width=1.01\linewidth]{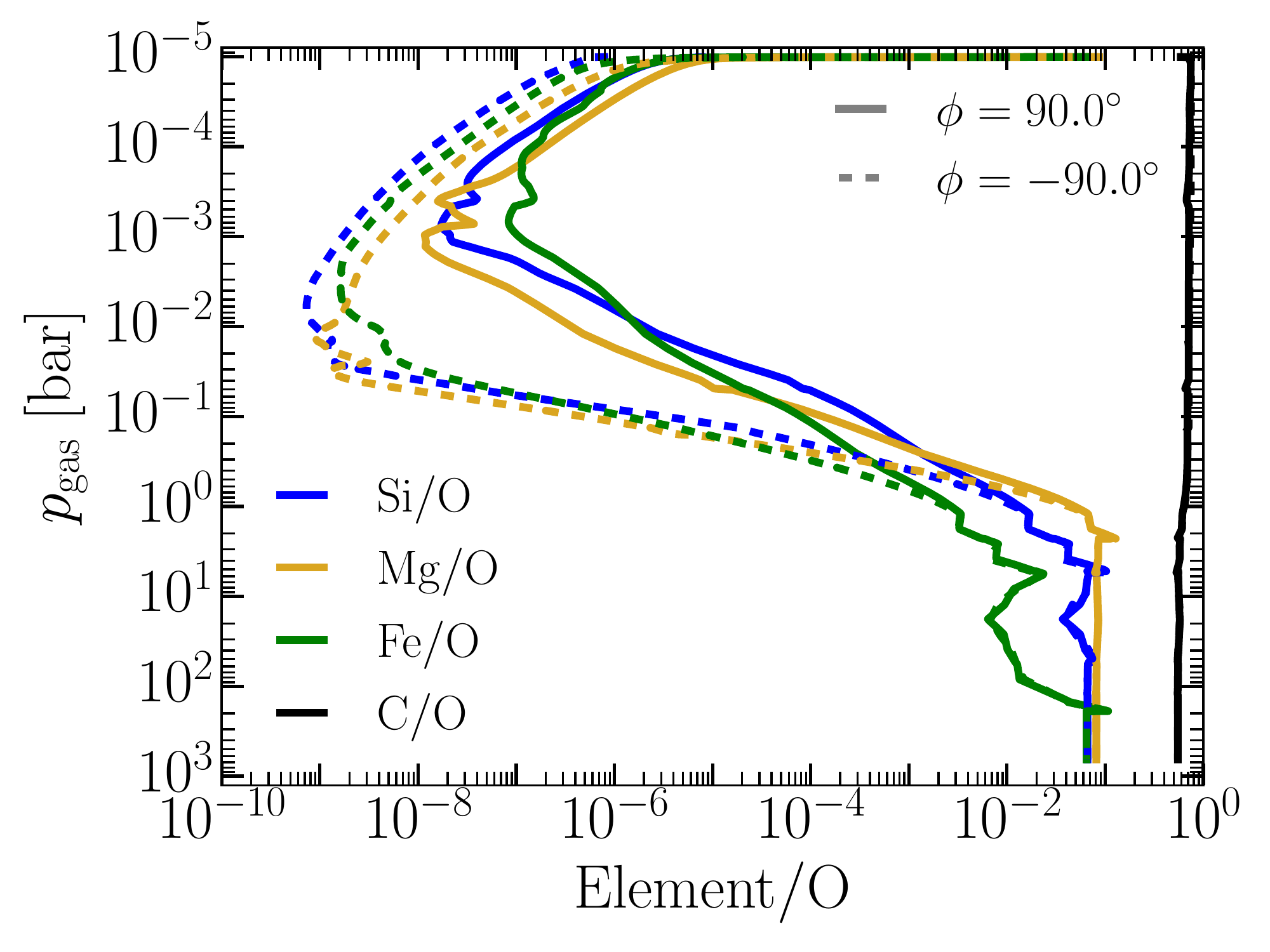}\\*[-0.3cm]
    \includegraphics[width=1.01\linewidth]{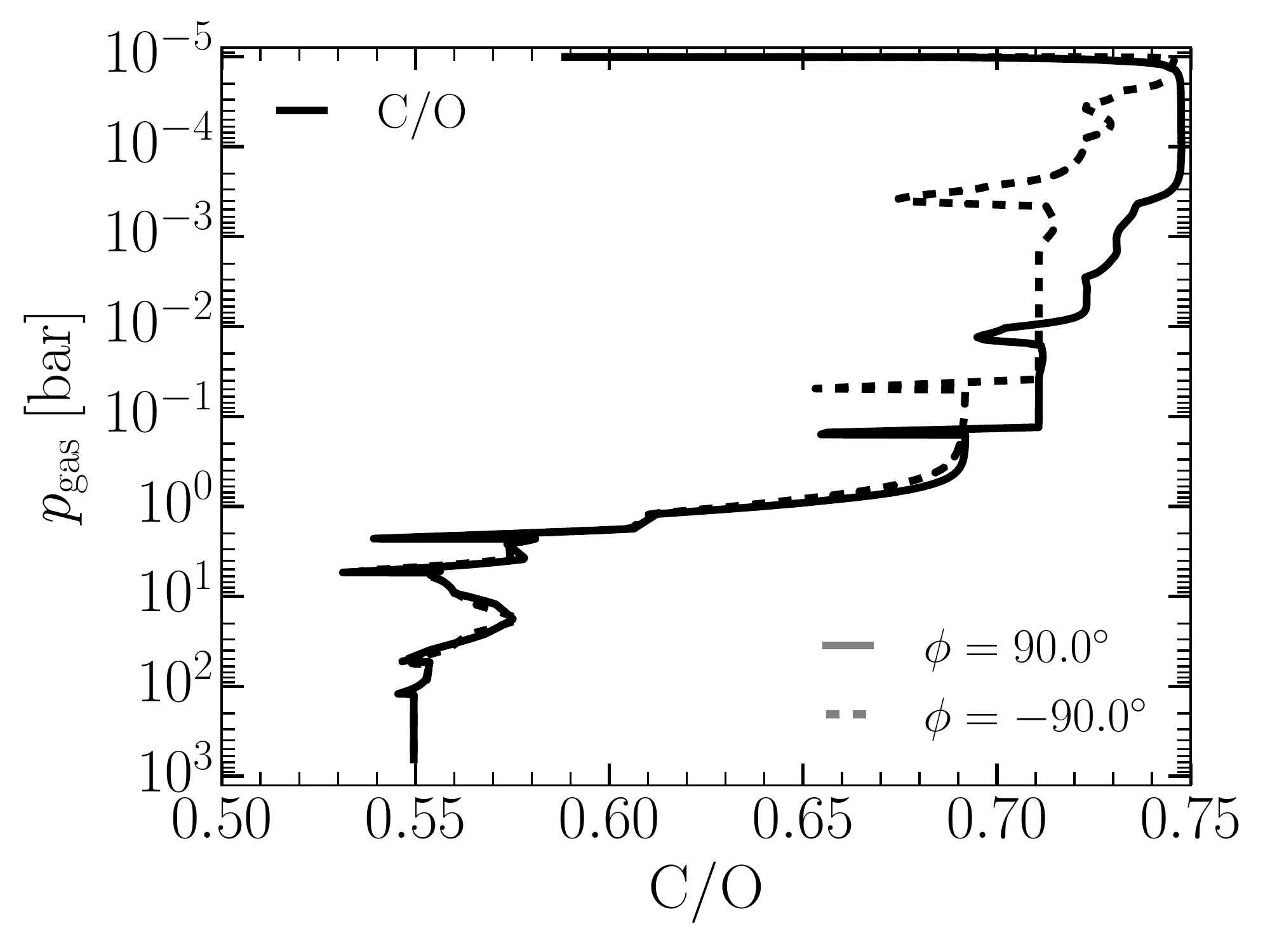}
    \caption{Gas phase abundance ratios, for morning ($\phi = -90.0$, dashed lines) and evening ($\phi = 90.0$, solid lines) terminators. Some morning/evening terminator differences exists. 
    {\bf Top:} The mineral ratios Si/O, Mg/O, Fe/O are strongly affected by element depletion due to cloud formation. {\bf Bottom:}
    C/O ratio appears lesser affected but oxygen is two orders of magnitude more abundant in a solar set of element abundances. C/O is therefore  shown on a linear scale, where C/O$=0.55$ is the Solar values.}
    \label{fig:Gas_Abunds}
\end{figure}

We do not consider carbon condensate species and have demonstrated the mineral CCN formation is more prominent in oxygen-rich atmospheres than the formation of hydrocarbon hazes (\citealt{2020A&A...641A.178H}).

The 2D terminator slice plots in Fig.~\ref{VsVtot} visualise the cloud particle material compositions in units of material volume fraction, $V_{\rm s}/V_{\rm tot}$ (Eqs. 25, 26 in \citealt{Helling2006} and Eq. 6 in \citealt{2008A&A...485..547H}). The 16 materials that are considered to contribute to the bulk growth of the cloud particles for our WASP-96b model are collected into four groups: metal oxides, silicates, high temperature condensates and salts (for details see caption of Fig.~\ref{VsVtot}). With a view to the possible observability of cloud particles through their spectral fingerprints, important results are:

\begin{itemize}
\item Cloud particles are made of a mix of all materials that are thermally stable at the local $(T_{\rm gas}, p_{\rm gas})$.
\item The silicate materials (\ce{MgSiO3}[s], \ce{Mg2SiO4}[s], \ce{Fe2SiO4}[s], \ce{CaSiO3}[s]) dominate almost everywhere the material composition of the cloud particles that form the clouds of WASP-96b. Their dominant contribution, however, varies between 40\% in the low-pressure regions ($p_{\rm gas}\lesssim10^{-1.5}\,$bar on the morning terminator, $p_{\rm gas}\lesssim10^{-2.5}\,$bar on the evening terminator) and >90\%.
\item The metal oxides (\ce{SiO}[s], \ce{SiO2}[s], \ce{MgO}[s], \ce{FeO}[s], \ce{Fe2O3}[s]) are the second most important materials reaching a contribution of no more than 40\% in the same locations where the silicates reach their 40\% level.
\item The high-temperature condensates (\ce{TiO2}[s], \ce{Fe}[s], \ce{FeS}[s], \ce{Al2O3}[s], \ce{CaTiO3}[s]) dominate the cloud particle's material composition at the cloud base where all silicates, metal oxides and salts have evaporated.
\item Salts (KCl[s], NaCl[s]) are negligible as bulk species.
\end{itemize}

We therefore suggest that the cloud particles in the low-pressure regions accessible by transmission spectroscopy should be dominated of a mix of $\approx$ 40\% metal oxides and 40\% silicates. These particles have a mean particle size of $\lesssim 10^{-2}\mu$m across the terminator region accessible by transmission spectroscopy.

\subsection{Mineral ratios and the mean molecular weight}

Mineral ratios (element ratios) like Si/O, Mg/O, Fe/O and the carbon-to-oxygen ratio, C/O, may help to link observations to evolutionary stages of exoplanets \citep{21KhMiDe,16ObEd}. The challenge with this inference is that observations have so far only provides snippets of the atmosphere spectrum, i.e. information about limited wavelength intervals. Consequently, inferences of element ratios often hinge on retrieved abundances of individual gas species. This situation, however, may change with the recent JWST observations, although the new JWST observations are still limited in range as they are only from NIRISS/SOSS.

The carbon abundance can only exceed the oxygen abundance if either carbon is produced (like in AGB stars) or inserted otherwise into the system (e.g. meteoritic influx), or if the oxygen is simply reduced below the original carbon-abundance level. The latter might occur if the planet already starts out with a primordial atmosphere that is oxygen-depleted due to the formation of \ce{H2O}/CO\, ices inside the planet-forming disk.  The accretion of the gaseous envelope from this oxygen-depleted gas and further processing of the atmospheric gas into clouds can then lead to a C/O>1 (\citealt{2014Life....4..142H}).

Compared to the the Si/O, Mg/O, and Fe/O ratios, C/O is not affected as strongly. The lower abundances of Si, Mg, and Fe are more significantly depleted by condensation compared to oxygen (Fig.~\ref{fig:Gas_Abunds}). However, the C/O ratio is enhanced in regions of cloud formation, due to oxygen element depletion from the gas phase through condensation (Fig.~\ref{fig:Gas_Abunds}, bottom).  The maximum value suggested by our simulations is C/O=0.75, which also occurs in the upper modelled atmosphere and thus likely in the optically thin atmosphere observed in transmission.
We note that also the set of Y and T-type brown dwarfs \cite{2022arXiv220504100G} appears to reach super-solar C/O. 

A decrease below the undepleted (solar) C/O value of 0.55 would suggest an increased amount of oxygen in the atmosphere or a substantially decrease of the carbon in the atmosphere.  \cite{2020A&A...635A..31M} show for the ultra-hot Jupiter HAT-P-7b that advection driven non-equilibrium may occur for H$_2$O but C/O would be determined by the oxygen depletion by cloud formation.

We note that in this section the local gas-phase C/O is discussed that includes the depletion of oxygen by cloud formation. The retrieval procedure in Section~\ref{subsec:retrieval_results} addresses the bulk C/O, which is indicative of the formation history of the planet. It is, however, demonstrated in Sect. ~\ref{subsec:discussion} that the retrieval suggests two solutions as best (or most probable) fit to the observed data.

The mean molecular weight remains that of an H$_2$-dominated gas until deep in the atmosphere because the upper atmosphere temperatures remain $>2000$K (see Fig.~\ref{fig:1dprofiles}).

\begin{figure*}
     \includegraphics[width=0.5\linewidth]{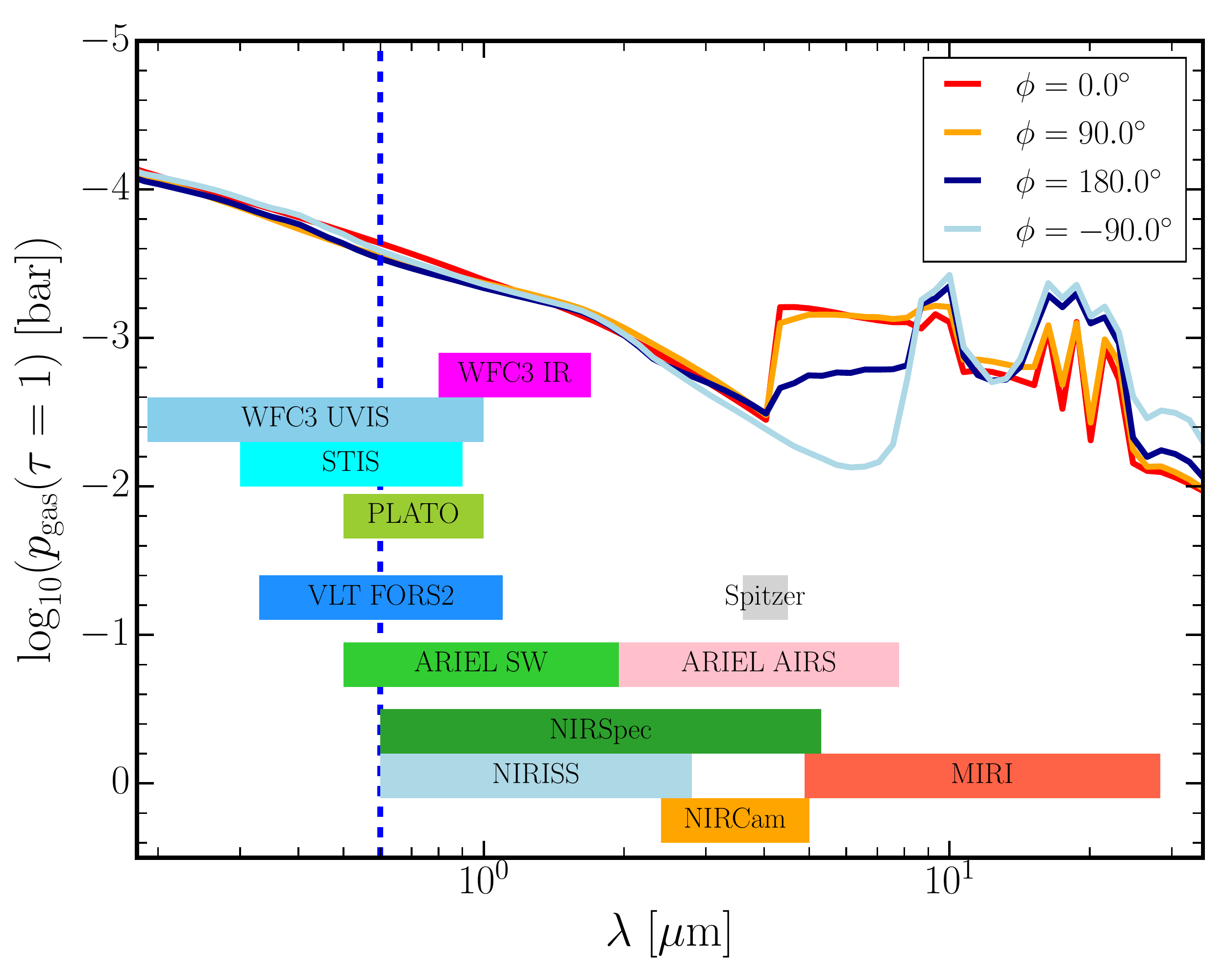}
     \includegraphics[width=0.5\linewidth]{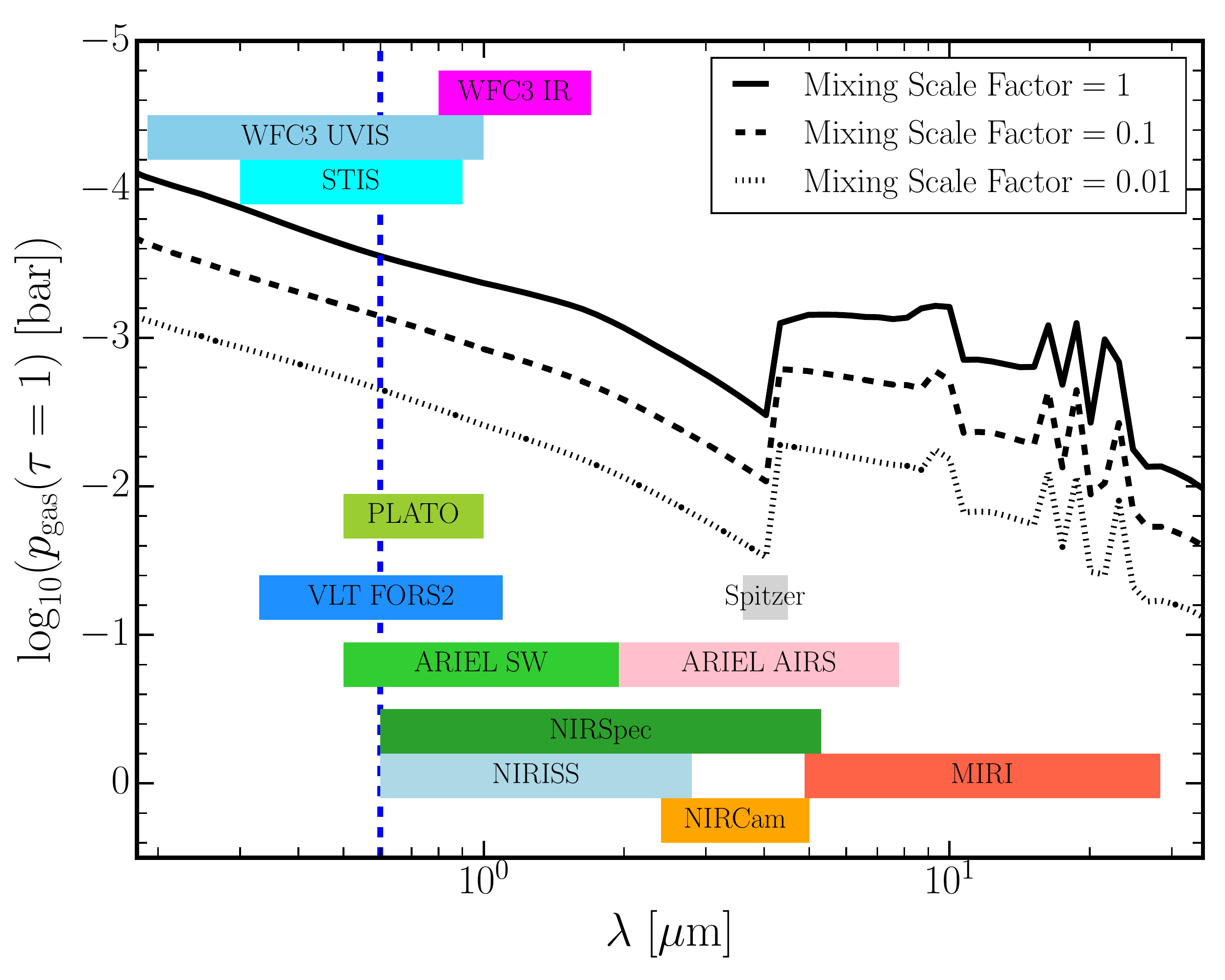}\\
     \includegraphics[width=0.5\linewidth]{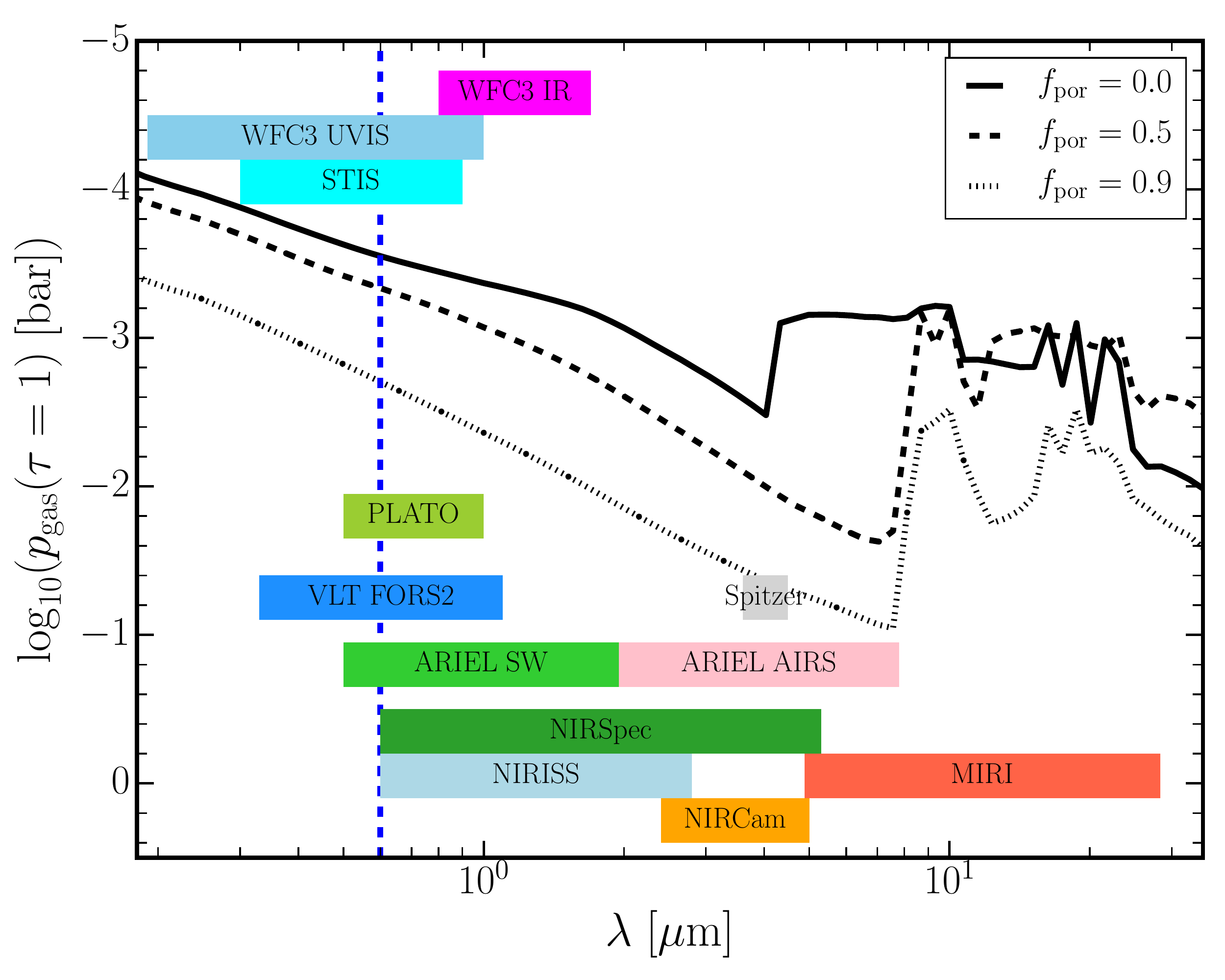}
     \includegraphics[width=0.5\linewidth]{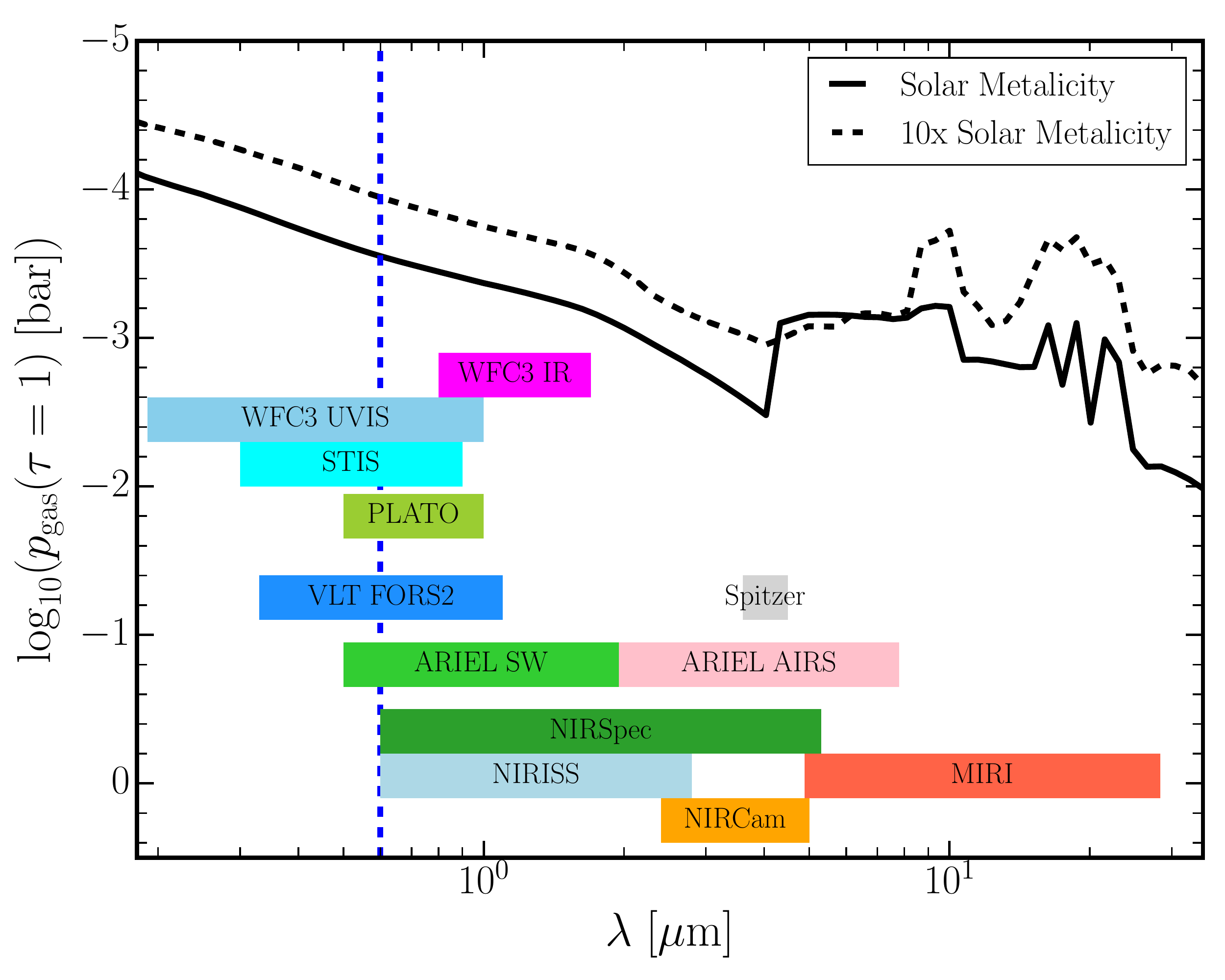}
    \caption{The changing slope of the optically thick pressure levels ($p_{\rm gas}(\tau(\lambda)=1)$ within optical to mid-IR wavelengths ($0.2-30\,\mu{\rm m}$). Blue dashed line indicates the location of the sodium line, coloured bars show wavelength ranges of various missions.  {\bf Top Left:} $\tau(\lambda)=1)$-slopes for the morning and evening terminator, the sub-stellar point ($\phi=0\degree$, red) and the anti-stellar point ($\phi=180\degree$, dark blue).  The remaining plots show parameter studies for the evening terminator ($\phi=90\degree,\ \theta=0\degree$). {\bf Top Right:} Different mixing efficiencies. {\bf Bottom Left:} The effect of cloud particle porosity. {\bf Bottom Right:} The effect of element abundances.}
    \label{fig:Optical_depth}
\end{figure*}

\subsection{Where clouds on WASP-96b get optically thick}
\label{subsec:Cloud_opt_depth}

Here we compute, based on our microphysical cloud model, the pressure where the optical depth of the clouds reaches unity ($p_{\rm gas}(\tau(\lambda)=1)$) - here after referred to as the `cloud deck'. This definition aligns with the cloud top pressure recovered in retrievals, although the exact physical interpretation of such parameters depends on the cloud model and observation geometry \citep{2020SSRv..216...82B}. Here we use the vertically integrated optical depth for individual 1D profiles, this neglects 3D effects. However, given the homogeneity of cloud formation in these models of WASP-96b, the effects of slant observing geometry can be corrected for through a correction factor \citep{Fortney2005}. Furthermore, the optical depths derived here depend on the computational domain of the GCM, as above this level we are unable to model cloud formation. Previously \cite{Helling2022} has shown that restricted GCM domains misses the `mineral haze layer' at higher altitudes. 

Taking four key points around the equator ($\theta=0.0^{\circ}$) of WASP-96b (Fig.~\ref{fig:Optical_depth} top left), we see that for wavelengths $< 4\mu{\rm m}$ indeed the cloud deck is consistent. Nonetheless, in this range the cloud deck still shows a substantial slope with wavelength. For example between $0.3...5\,\mu{\rm m}$ for the morning terminator, the cloud deck pressure varies between $1.3\times 10^{-4}...6\times 10^{-3}\,{\rm bar}$. This suggests flat cloud decks are a poor fit across this wavelength range, which becomes more important as we move towards an era of broader wavelength coverage with a single mission. In the mid-IR, it is a different story, spectral differences may be expected within the silicate feature $\lambda$-range, visible with NIRSPec, NIRCam, ARIEL and MIRI. Spitzer sits right on this silicate edge at $\sim4\,\mu{\rm m}$, which may complicate combining datasets. In particular focusing on the two terminator limbs there is substantial differences in the cloud deck from $4...8\,\mu{\rm m}$ of nearly an order of magnitude in pressure. 

The high pressure cloud decks (or cloud-free atmospheres) previously retrieved by \cite{2018Natur.557..526N} in order to fit the Na line (dashed blue vertical line) are inconsistent with microphysical cloud formation. At the Na line, the model produces a cloud deck at $\sim2.5\times 10^{-4}\,{\rm bar}$, whereas pressures as high as $10^{-2}\,{\rm bar}$ are needed to fit the pressure broadening observed. However, given the pressure-temperature structures shown in Fig.~\ref{fig:1dprofiles}, simple thermal stability arguments will expect clouds to form. Retrievals cannot simply infer condensable-favouring global temperatures and claim low cloud decks, or cloud-free atmospheres without proposing some mechanism for cloud suppression.

We therefore explore 3 parameters, vertical mixing efficiency, cloud particle porosity, and metallicity as to how they impact the cloud deck pressure level. Reduction of the vertical mixing efficiencies is achieved by increasing the mixing timescale. This is done by the introduction of a `mixing scale factor' $f_{\rm mix}$, such that the new mixing timescale is

\begin{equation}
    \tau_{\rm mix}^{\rm eff} = \frac{1}{f_{\rm mix}}\tau_{\rm mix} 
\end{equation}

where $\tau_{\rm mix}$ is the mixing timescale from the GCM derived the same way as in \cite{Helling2022}. Alternatively, using $K_{\rm zz} = H_{\rm p}\tau_{\rm mix}$ implies that similarly $K_{zz}$ is reduced by factors of $f_{\rm mix}$. Reduction of the vertical mixing efficiencies ($f_{\rm mix} = 0.1,\,0.01$) reduce the cloud deck height by an order of magnitude in pressure, but otherwise maintains the wavelength dependence. The mechanism for such inefficient mixing requires further investigations.

Porosity of cloud particles is included as described in \cite{20SaHeMi}, reducing the  effective material density of cloud particles $\rho_{\rm s}^{\rm eff}$ through the porosity factor $f_{\rm por}$, $\rho_{\rm s}^{\rm eff} = \rho_{\rm s}(1-f_{\rm por})$.

Mass conservation ensures that correspondingly the radius of cloud particles are also increased. In addition, the porosity of the cloud particle is incorporated into opacity calculation through inclusion of vacuum refractive index into effective medium theory \citep{Bruggeman1935,Looyenga1965}, in proportion to the porosity factor of the particle.

Highly porous particles reduce the opacity of the clouds, even with enhanced cloud mass load and a higher onset of bulk growth, as was the case in \cite{20SaHeMi}, although the effect here is substantially larger than in the hotter atmosphere ($T_{\rm eff}=1800\,{\rm K}$) presented there. Unlike reductions in mixing efficiency, increased cloud particle porosity affects wavelengths $>4\,\mu{\rm m}$ non-uniformly, the cloud deck pressure level at the silicate feature wavelengths is particularly affected by compact vs porous cloud particles.

The porosity and shape of cloud particles (or more generally aerosols) in exoplanet atmospheres remains uncertain \citep{Gao2021_Rev}. However, there is some tentative evidence of porous cloud particles (generated through collisions) as an explanation for the flat spectrum of the mini-Neptune GJ-1214b \citep{Ohno2020}. Although, as shown in \cite{Samra2022}, collisions in exoplanet atmospheres when including turbulence are expected to be destructive rather than producing larger aggregate particles. However, hazes of fractal aggregates have been inferred for a number of bodies, both Solar system and exoplanets (e.g. \citealt{Adams2019}). Porosity has also been examined in the other astrophysical contexts, such as protoplanetary disc dust~\citep{Woitke2016}, albeit only with a factor of 25\% vacuum assumed. Overall whether cloud particles could reach such high (90\%) porosities is unknown, but it still remains an effect which is important to consider in interpreting exoplanet observations.

\begin{table*}[!tp]
	\caption{Parameters and associated priors for the retrieval of WASP-96b VLT/HST/Spitzer data using ARCiS. A description of these parameters can be found in \cite{20MiOrCh.arcis,22ChMi}.}
	\label{t:ret_pars} 
	\centering  
	\begin{tabular}{lcc}
		\hline
		\hline
		\rule{0pt}{3ex}Name & Description & Priors \\
		\hline
		\multicolumn{3}{l}{\rule{0pt}{3ex}\textbf{Parameters included for both cloud formation and clear retrievals}}\\
		\rule{0pt}{3ex}$R_{\rm p}$ &  \makecell{\rule{0pt}{3ex} Planet radius} & \makecell{\rule{0pt}{3ex} 5~$\sigma$ around lit. value$^a$, units of $R_J$, flat linear prior}\\
		\rule{0pt}{3ex}log$_{10}$($g_{\rm p}$) &  \makecell{\rule{0pt}{3ex} Base-10 logarithm of the planet surface gravity$^b$} & \makecell{\rule{0pt}{3ex} 1 to 5 (with $g$ given in cgs units)}\\[2mm]
				\rule{0pt}{3ex}C/O &  \makecell{\rule{0pt}{3ex} Atmospheric carbon-to-oxygen ratio} & \makecell{\rule{0pt}{3ex} 0.1 (sub-solar) to 1.3 (super-solar), flat linear prior}\\
						\rule{0pt}{3ex}N/O &  \makecell{\rule{0pt}{3ex} Atmospheric nitrogen-to-oxygen ratio} & \makecell{\rule{0pt}{3ex} 0 (sub-solar) to 0.3 (super-solar), flat linear prior}\\
		\rule{0pt}{3ex}Si/O &  \makecell{\rule{0pt}{3ex} Atmospheric silicon-to-oxygen ratio} & \makecell{\rule{0pt}{3ex} 0 (sub-solar) to 0.3 (super-solar), flat linear prior}\\
						\rule{0pt}{3ex}[Z] &  \makecell{\rule{0pt}{3ex} Atmospheric metallicity (global) } & \makecell{\rule{0pt}{3ex} -3 (sub-solar) to 3 (super-solar) dex, flat linear prior}\\
						\rule{0pt}{3ex}log$_{10}$($\gamma$)&  \makecell{\rule{0pt}{3ex} Ratio of visible to IR opacity} & \makecell{\rule{0pt}{3ex} -2 to 2, flat log prior}\\
		\rule{0pt}{3ex}f$_{\rm irr}$&  \makecell{\rule{0pt}{3ex} Irradiation parameter} & \makecell{\rule{0pt}{3ex} 0 to 0.25, flat linear prior}\\
		\rule{0pt}{3ex}log$_{10}$($\kappa_{\rm IR}$)&  \makecell{\rule{0pt}{3ex} Infrared opacity} & \makecell{\rule{0pt}{3ex} -4 to 4 cm$^2$~g$^{-1}$, flat log prior}\\
			\rule{0pt}{3ex}T$_{\rm int}$ &  \makecell{\rule{0pt}{3ex} Temperature at an optical depth $\tau$~=~$\frac{2}{3}$ as caused by \\ internal heat from the planet} & \makecell{\rule{0pt}{3ex} 10 to 3000~K, flat log prior}\\[2mm]
		\multicolumn{3}{l}{\rule{0pt}{3ex}\textbf{Parameters included in cloud formation retrieval only}}\\
		\Xhline{\arrayrulewidth}
		\rule{0pt}{3ex}$K_{\rm zz}$ &  \makecell{\rule{0pt}{3ex} Cloud diffusion coefficient} & \makecell{\rule{0pt}{3ex} 10$^{5}$ - 10$^{12}$~cm$^2$~s$^{-1}$, flat log prior}\\
		\rule{0pt}{3ex}$\log_{10}\dot\Sigma$ &  \makecell{\rule{0pt}{3ex} Nucleation rate} & \makecell{\rule{0pt}{3ex} 10$^{-17}$ - 10$^{-7}$~g~cm$^{-2}$~s$^{-1}$, flat log prior}\\
		\hline
		\hline
	\end{tabular}
\flushleft{$^a$: \cite{14HeAnCa}
\flushleft{$^b$: We use a gaussian prior on the mass based on radial velocity measurements of WASP-96b~\citep{14HeAnCa}}. We compute the mass based on $R_p$ and log($g_p$) and then place the prior on this derived parameter.}
\end{table*}

For this work we have so far assumed solar elemental abundances \citep{2009ARA&A..47..481A} for the gas composition of the atmosphere. \cite{2022MNRAS.tmp.1485N} derive elemental abundances for the host star WASP-96 using FEROS spectra. These abundances are enhanced with respect to the solar abundances of \citep{2009ARA&A..47..481A}, at most around a factor of $\sim 3x$ solar (for example Mg and Mn - see their Table 4). However, it is unclear if stellar metallicities correlate with planetary atmosphere abundances \citep{Teske2019}. Nonetheless, we examine the effect of enhanced metallicity for cloud formation in the atmosphere using a 10x Solar metallicity for the planetary atmosphere. This is done by increasing abundance of everything except H and He and re-normalising to H abundance. Metallicity for WASP-96b, based on formation arguments for lower mass gas-giant planets (e.g. \citealt{Carone2021,Chachan2019, Schneider2021b}), is expected to be super-stellar. The enhanced metallicity increases cloud formation and leads to a higher cloud deck (bottom right of Fig.~\ref{fig:Optical_depth}), making it even harder to reconcile with super-solar retrievals.

\section{Clouds affecting  $0.3 - 5\mu$m spectra}

Solving the radiative transfer problem through a prescribed atmospheric structure with clouds allows us to explore the effect of cloud properties on the possible spectral appearance of WASP-96b within the wavelength range that has so far been explored with VLT/FORS2, HST/WFC3 and Spitzer. This procedure is known as forward modelling within the retrieval community. The ARCiS~\citep{20MiOrCh.arcis,18OrMi.arcis} frame work is applied here. We further investigate the possibility of finding a solution to the VLT/FORS2, HST/WFC3 and Spitzer data including the effect of clouds using a retrieval setup which includes a parameterised cloud formation routine.

\smallskip
Retrieval studies of the HST/WFC3 and Spitzer/IRAC transmission spectra of WASP-43b also previously struggled to find evidence for the presence of clouds in the atmospheric spectra~\citep{2014ApJ...793L..27K,kchubb2019}, at least at the pressure levels probed by these observations. However works like \cite{2020A&A...641A.178H} studied the formation of mineral clouds and hydrocarbon hazes in the atmosphere of WASP-43b and demonstrate that it is rather unlikely that WASP-43b is cloud-free. \cite{2022ApJ...930...93R} performed a detailed study of the effects of clouds on emission phase curve spectra of WASP43b by computing various forward models for clear and cloudy atmospheres using PICASO~\citep{19BaFrLu_web,20BaRo_web}.

They found that cloudy phase curves, produced from 3D models, provide much better agreement with the WFC3 and Spitzer nightside data than cloud-free models.

\subsection{Retrieval results}
\label{subsec:retrieval_results}

 \begin{figure*}
  \hspace*{-0.5cm}  
   \includegraphics[width=0.53\linewidth]{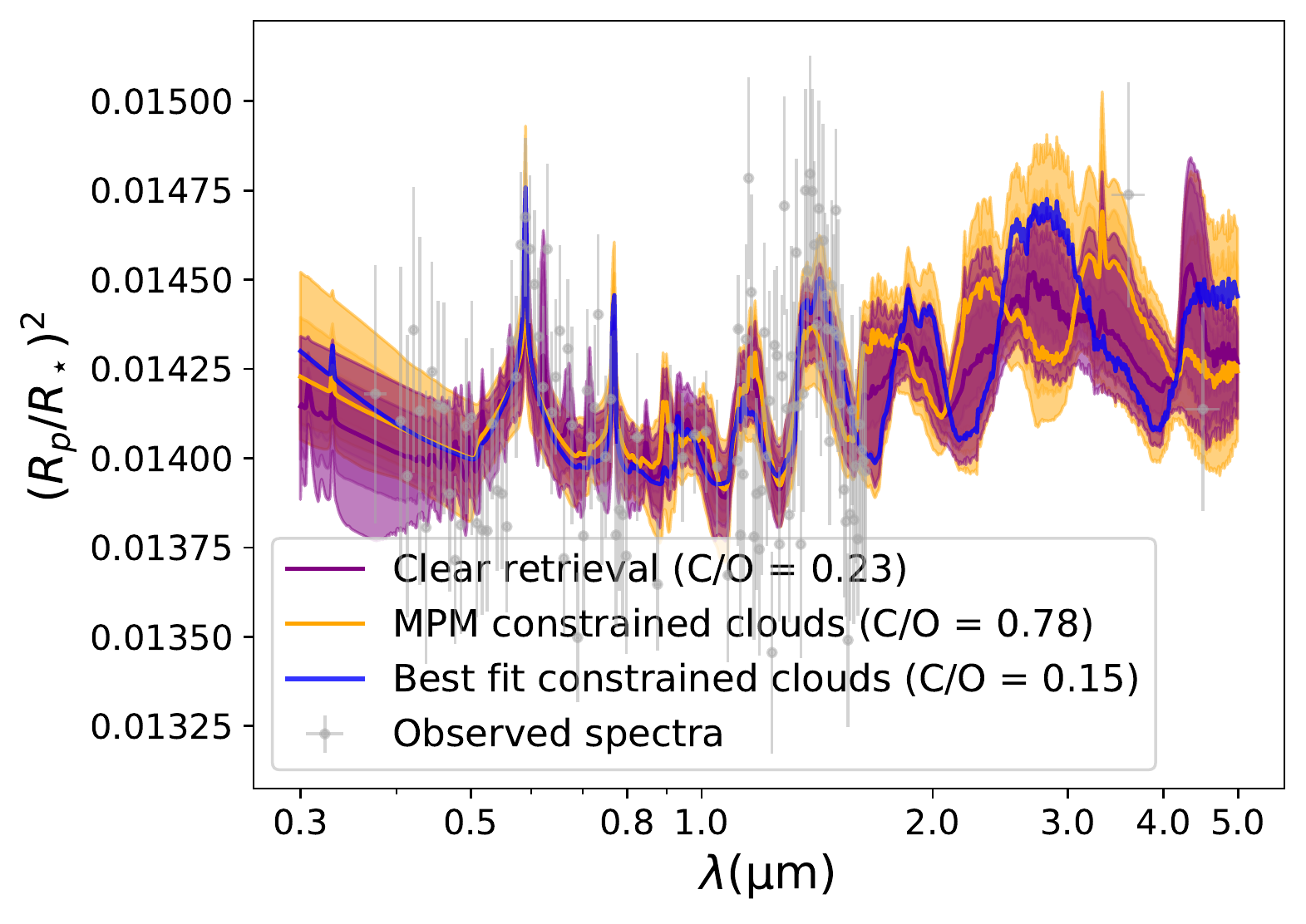}
    \includegraphics[width=0.53\linewidth]{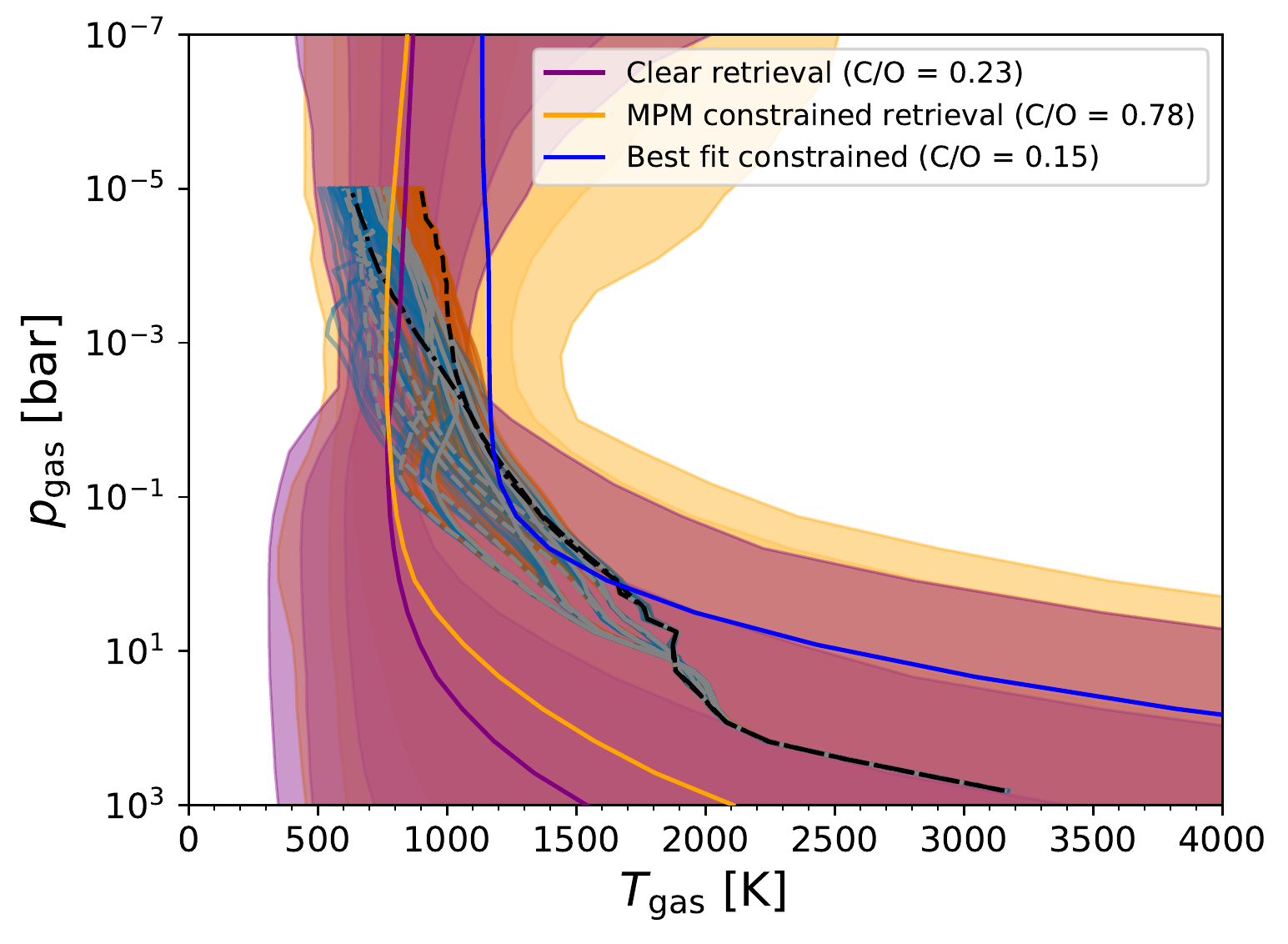}
    \caption{The best fit models (1, 2 and 3 $\sigma$ bounds shading) retrieved with ARCiS from the \cite{2022MNRAS.tmp.1485N} pre-JWST observations. In purple is the ARCiS retrieval with no clouds included, and in orange with cloud formation included (as described in \citealt{18OrMi.arcis}). Both retrievals assume equilibrium chemistry. In the cloud formation case the retrieved parameters are shown in Figure~\ref{fig:corner_cloud_formation}, and for the no cloud case they are shown in Figure~\ref{fig:corner_no_clouds}. MPM refers to the ‘Median Probability Model’, which is slightly different to the `best fit' model; see text. {\bf Left:} Retrieved spectrum, {\bf Right:} the retrieved pressure-temperature profile, for reference the GCM temperature-pressure structures are over-plotted, using the same colour scheme as Fig.~\ref{fig:1dprofiles}.}
    \label{fig:comp_cloud_formation_no_cloud_ret}
\end{figure*}

Here, we present the results of our retrievals of the observed VLT~\citep{2018Natur.557..526N}, HST, and Spitzer~\citep{2022MNRAS.tmp.1485N} transit spectra using atmospheric retrieval and modelling code ARCiS~\citep{20MiOrCh.arcis,18OrMi.arcis}. Molecular and atomic opacities in the form of k-tables are utilised from various sources such as ExoMol~\citep{ExoMol2020,jt631}, HITEMP~\citep{HITEMP}, HITRAN~\citep{HITRAN2020}, MoLLIST~\citep{MOLLIST}, and NIST~\cite{NISTWebsite}, as detailed in \cite{20ChRoAl.exo}. We include the following set of species (and opacities computed by the listed line list) in both sets of retrievals, which compute spectra under the assumption of equilibrium chemistry using GGchem~\citep{2018A&A...614A...1W}:
H$_2$O~\citep{jt734}, CO$_2$~\citep{20YuMeFr.co2}, CH$_4$~\citep{17YuAmTe.CH4}, CO~\citep{15LiGoRo.CO},
        OH~\citep{MOLLIST,18YoBeHo.OH},
        AlO~\citep{ExoMol_AlO},
        K~\citep{16AlSpKi.broad,19AlSpLe.broad,NISTWebsite},
        Na~\citep{19AlSpLe.broad,NISTWebsite},
        NH$_3$~\citep{ExoMol_NH3},
        TiO~\citep{ExoMol_TiO},
        VO~\citep{ExoMol_VO},
        H$_2$S~\citep{jt640},
        HCN~\citep{jt570},
        PH$_3$~\citep{jt592},
        C$_2$H$_2$~\citep{ExoMol_C2H2},
        CN~\citep{14BrRaWe.CN}, and
        SiO~\citep{jt563,21YuTeSy}.

We ran two different retrievals: A constrained cloud formation retrieval, following the method of \cite{20MiOrCh.arcis}, on the combined HST/VLT/Spitzer data set (including HST/VLT offset) of \cite{2022MNRAS.tmp.1485N}, and a completely clear (no parameters describing clouds included) retrieval, for comparison. Both retrievals assume chemical equilibrium for the gas-phase chemistry, which is computed after the cloud formation computations. The parameters retrieved and their priors can be found in Table~\ref{t:ret_pars}.  The best fit retrieved spectra, with 1, 2, and 3~$\sigma$ shading, can be found in Figure~\ref{fig:comp_cloud_formation_no_cloud_ret} (left).
The observed spectra is plotted for comparison. Our retrieved posterior distributions can be found in Figure~\ref{fig:corner_cloud_formation} for the cloud formation case, and Figure~\ref{fig:corner_no_clouds} for the clear atmosphere case. The first four parameters in Figures~\ref{fig:corner_cloud_formation}~and~\ref{fig:corner_no_clouds} describe the pressure-temperature parametrisation of \cite{10Guillot.exo}. These are used to construct the pressure-temperature profiles for the two retrievals given in Figure~\ref{fig:comp_cloud_formation_no_cloud_ret} (right). The pressure-temperature profiles from the GCM used in our atmospheric modelling and cloud formation models are plotted for comparison. These can also be compared to Figure~5 of \cite{2022MNRAS.tmp.1485N}, with the clear retrieval pressure-temperature profile in our Figure~\ref{fig:comp_cloud_formation_no_cloud_ret} (right) appearing very similar to their retrieved pressure-temperature profile, which also assumes equilibrium chemistry.

The two panels of Figure~\ref{fig:comp_cloud_formation_no_cloud_ret} illustrate solutions from the cloud formation retrieval based on slightly different model outputs. The `best fit' model is the one with the lowest $\chi^2$ when compared to the observations. However, there are other possible solutions which can also fit the data similarly well. The `Median Probability Model' (MPM)~\citep{BaBe2004,21BaBeGe} is based on the posterior distributions shown in Figure~\ref{fig:corner_cloud_formation}, and represents an alternative region of parameter space for the cloud formation retrieval presented in this work. In particular, the posterior distribution for the retrieved C/O has two potential solutions; one with high C/O ($\sim$0.78; the `MPM' solution), and one with lower C/O ($\sim$0.15; the `best fit' solution). It can be seen from the left panel of Figure~\ref{fig:comp_cloud_formation_no_cloud_ret} that the retrieved best fit and MPM spectra are very similar in the region covered by the VLT and HST observations, and fit the observed data similarly well. They deviate, however, in the region of the spectra for wavelengths longer than the HST data. The same is true for the clear retrieval; it deviates most in the region not covered by the presented observed data. It is therefore difficult to distinguish them based on the presented observed data. The bottom two panels of Figure~\ref{fig:cloud_abundances} give the molecular abundances (volume mixing ratios) as a function of pressure for the MPM (left) and best-fit (right) solutions. There are large variations in the CO and CH$_4$ abundances in particular between these two solutions. The primary spectral features of both these molecules are not in the region of the spectra covered by the HST and VLT observations; around 2.3 and 5~$\mu$m for CO, and around 3.3~$\mu$m for CH$_4$. As illustrated by \cite{22GaMiCh}, other absorption features of CH$_4$ which are present in the region of the HST data are often masked by stronger H$_2$O features. Figure~\ref{fig:mixrat_GCM} shows the volume mixing ratios of various molecules predicted to be present at the equator of the morning (left) and evening (right) terminators, based on the GCM and kinetic cloud models of this work. CO is expected to be abundant in both cases, along with H$_2$O at most altitudes. These are more similar to the lower C/O retrieval solution in the bottom left panel of Figure~\ref{fig:cloud_abundances}. It can also been seen from Figure~\ref{fig:cloud_abundances} that the difference in H$_2$O abundance between the two retrieval solutions can have a noticeable effect on the spectra, again in particular in the region above 1.7~$\mu$m. This highlights the benefits of the JWST observations of WASP-96b~\citep{22PoBlBr_arxiv}, which do cover this region.

It can be seen by the right-hand panel of Figure~\ref{fig:comp_cloud_formation_no_cloud_ret} that the cloud formation retrieval solution with the lower C/O has a higher atmospheric temperature than the solution with the higher C/O. This is due to constraints placed by the cloud formation scheme. The same does not apply to the clear atmosphere retrieval; there is no constraint on the temperature based on what types of clouds are expected to form and a lower temperature is therefore preferred.
The cloud formation model with the lower C/O necessitates a higher atmospheric temperature at the pressure layers being probed by the observations. This is because a low C/O and therefore an abundance of oxygen would mean that K- and Na-silicate clouds are expected to form if the temperature is low enough. However, the strong gas-phase Na and K spectral features suggest that these species of clouds either have not formed efficiently in this atmosphere, at least not at a very large abundance to take all the Na and K away from the gas-phase, or they form but they are optically thin enough such that the gas-phase Na and K is still observable. 
A recent observation of planetary mass companion VHS~1256-1257~b using JWST's NIRSpec IFU and MIRI MRS modes revealed solid-states silicate features in the longer wavelength part of the atmospheric spectrum~\citep{22MiBiPa}. They conclude this as strong evidence for small silicate particles in the brown dwarf's atmosphere; as demonstrated by works such as \cite{04MiDoWa}, small particles lead to less overall extinction across the spectra, but a more prominent silicate feature. The atmospheric mixing is described by a $K_{\rm zz}=10^8\,\ldots\,10^9$ cm$^2$/s.  These observations suggest that silicate clouds do form in sub-stellar hot atmospheres and it is possible to directly observe their spectral signatures with JWST. 

This is either due to a lower C/O and therefore less oxygen present for these clouds to form from, or a lower C/O but a high enough atmospheric temperature to stop them from forming in abundance. We also ran a retrieval with cloud formation, but without the cloud species which condense Na and K included. In this case, the lower C/O solution was preferred, as there were no constraints placed based on the temperature of formation for these types of clouds. It should be noted that the C/O for the cloud formation retrievals is the bulk C/O, i.e. before clouds are allowed to form. Clouds will generally deplete oxygen from the atmosphere, so the retrieved C/O without cloud formation constraints (but where clouds should be present based on atmospheric temperature, such as for the clear retrieval), will be expected to be retrieved higher than the bulk C/O.
The models with the lowest C/O ratio show significant H$_2$O features in the region not covered by the HST and VLT data, which is not the case for the high C/O case. Since the newly observed JWST ERO data appears to show large H$_2$O spectral signatures~\citep{22PoBlBr_arxiv}, a situation with a lower C/O and higher atmospheric temperature could be preferred. A more detailed study of this JWST spectra of WASP-96b in the near-future should be very informative.

The reduced-$\chi^2$ value was very similar for the clear and the cloud formation retrievals (1.08 and 1.09, respectively), indicating that both can fit the observed data equally well. The cloud formation retrieval is however not considered statistically significant in comparison to the clear retrieval, i.e. the inclusion of extra parameters to characterise the clouds is not justified by a significant increase in how well the model fits the data. This has previously led to the conclusion that the atmosphere must be clear. Based on expectations from detailed simulations of cloud formation in the atmosphere of WASP-96b, as demonstrated in this paper, it would be very surprising from a physical point of view if the atmosphere is completely clear. We therefore try to offer some explanations to confront this seeming contradiction between what models predict (a cloudy atmosphere) and what the observed data appears to show (a clear atmosphere).

\begin{figure}
    \centering
     \includegraphics[width=\linewidth]{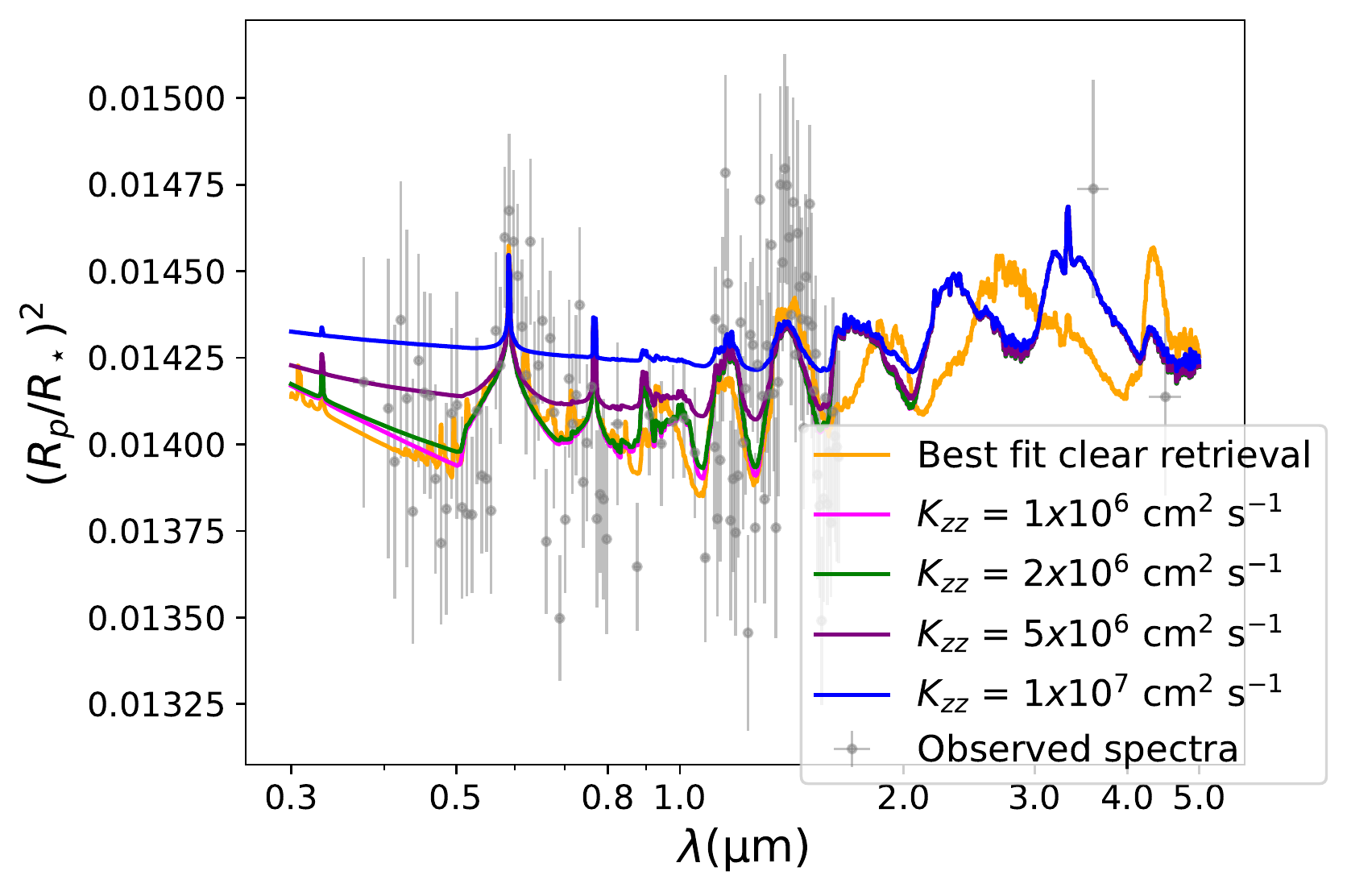}
    \includegraphics[width=\linewidth]{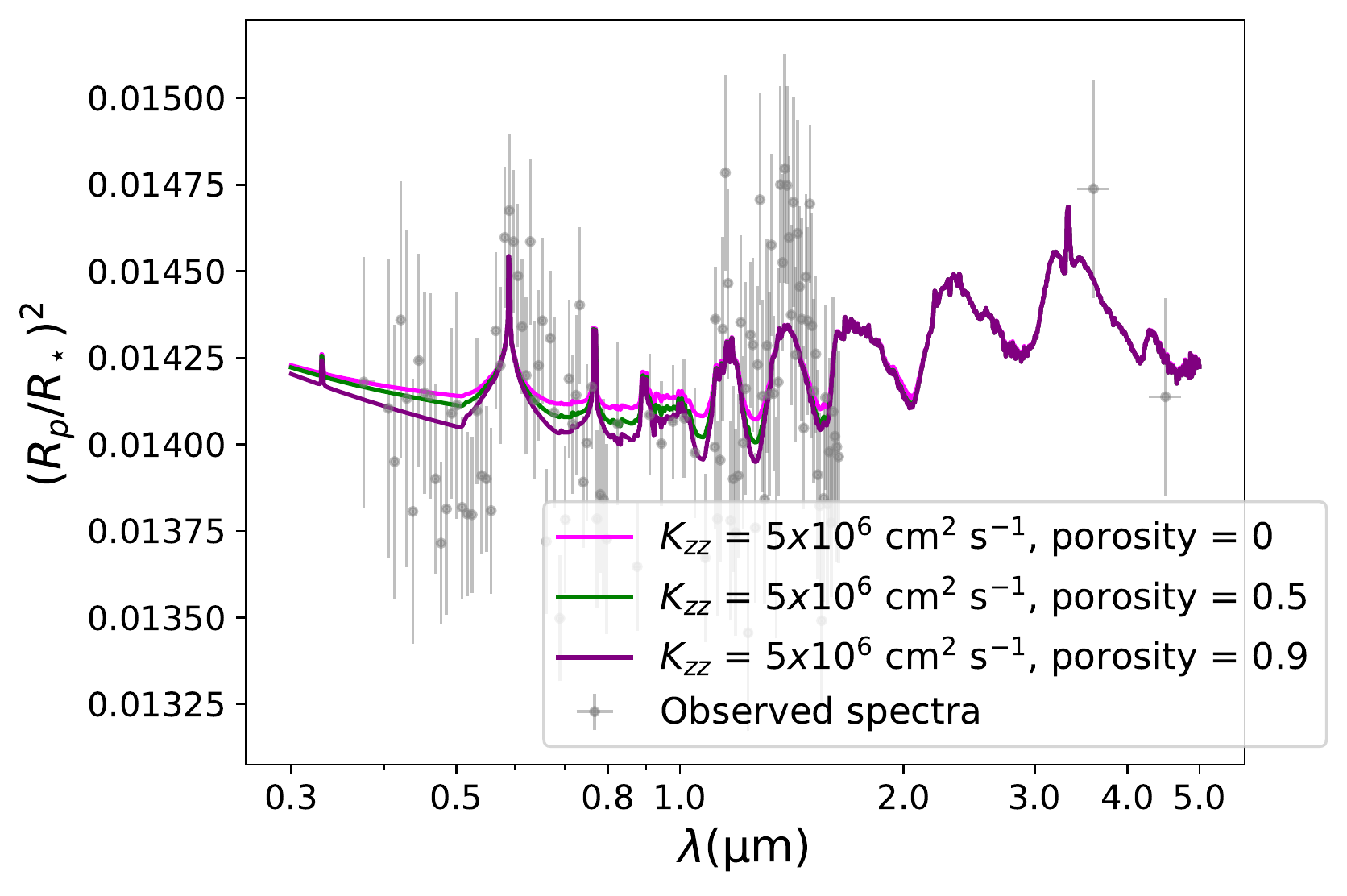}
    \caption{The best fit synthetic spectrum from the constrained retrieval of the \cite{2022MNRAS.tmp.1485N} data (VLT, HST, and Spitzer) with varying values of cloud diffusivity, $K_{zz}$, (top), and for varying cloud particle porosity for atmospheric models with cloud diffusivity $K_{zz}$~=~5$\times$10$^6$~cm$^2$~s$^{-1}$ (bottom).}
    \label{fig:cloud_formation_Kzz}
\end{figure}

\begin{figure}
    \centering
      \includegraphics[width=\linewidth]{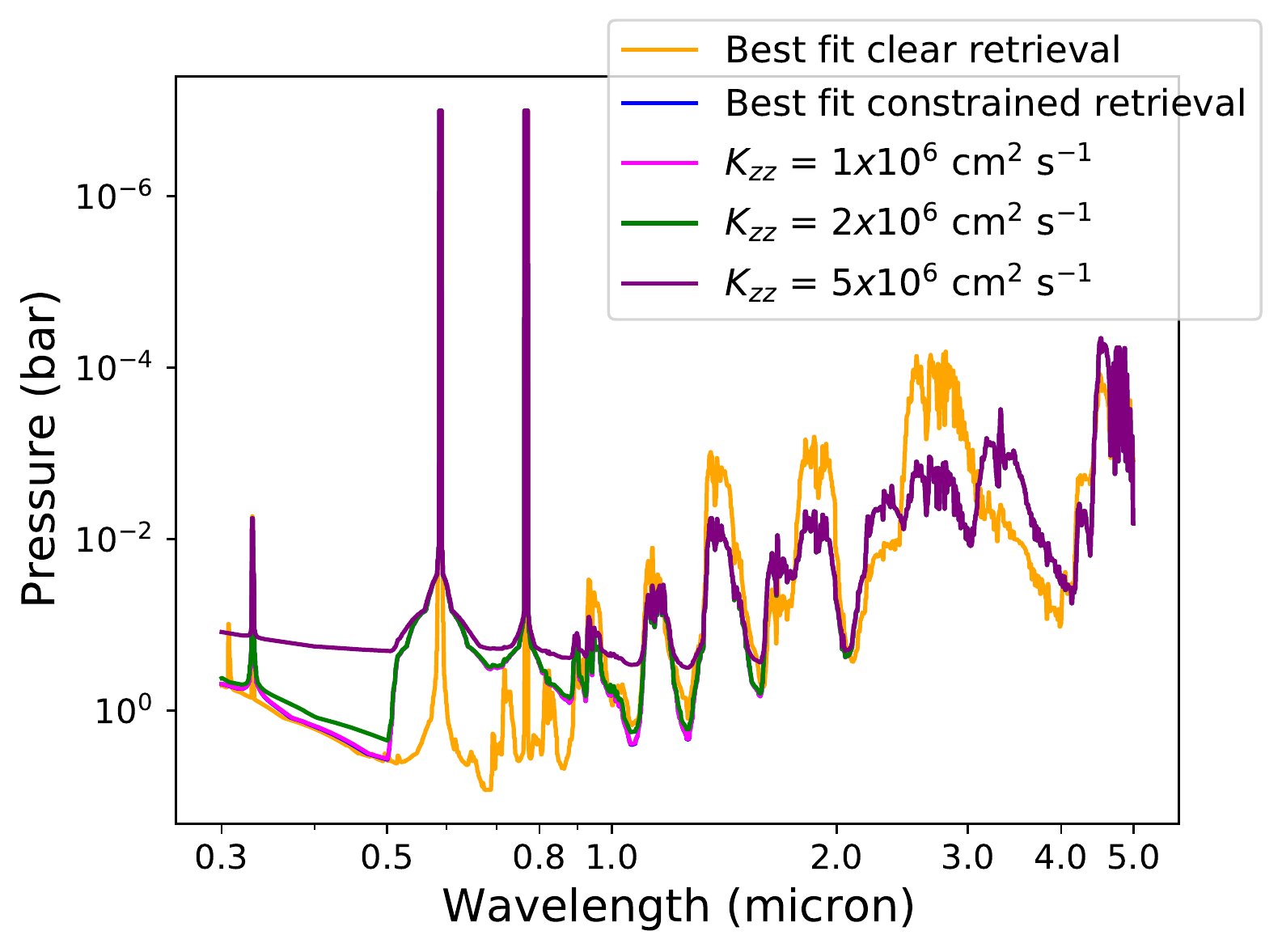}
    \caption{The same cloud formation setups with varying cloud diffusivity $K_{zz}$ as for Figure~\ref{fig:cloud_formation_Kzz}, but this time showing the pressure levels where optical depth $\tau$ reaches unity, as a function of wavelength. The sharp drop of the Na lines (green line) is artificial and caused by the cut-off in the opacity data available. }
    \label{fig:Kzz_pressure}
\end{figure}

Firstly, in Figure~\ref{fig:corner_cloud_formation}, the cloud diffusivity $K_{zz}$ (which represents the vertical mixing strength of cloud particles) is constrained to be no higher than around 2$\times$10$^6$~cm$^2$~s$^{-1}$. This corresponds to the reduced mixing scenario explored in Figure~\ref{fig:Optical_depth} (top right). A reduction in vertical mixing of cloud particles settles the cloud layer further down in the atmosphere, thus reducing its optical effects on blocking or muting gas-phase molecular or atomic signatures. We further illustrate this by computing a forward model using our best fit cloud formation retrieval, with varying values of $K_{zz}$ (Figure~\ref{fig:cloud_formation_Kzz}). Figure~\ref{fig:Kzz_pressure} shows the pressure level where the optical depth of the atmosphere reaches unity (i.e. it becomes optically thick) as a function of pressure for the models. It can be seen from Figure~\ref{fig:cloud_formation_Kzz} that lower values of $K_{zz}$, around 2$\times$10$^6$~cm$^2$~s$^{-1}$, allow for a clear detection of the Na doublet at around 0.6~$\mu$m, with the atomic line broadening by H$_2$ and He clearly visible. Increasing $K_{zz}$ mutes this, and other gas-phase, spectral features.
 Our retrieved value of vertical cloud diffusion parameter $K_{\rm zz}$ is between 1.7~$\times$~10$^5$ and 1.8~$\times$~10$^6$~cm$^2$~s$^{-1}$, within 1~$\sigma$ bounds, and between 2.1~$\times$~10$^4$	and 2.5~$\times$~10$^7$~cm$^2$~s$^{-1}$ within 3~$\sigma$ bounds. The priors on $K_{\rm zz}$ were $10^5...10^{12}$~cm$^2$~s$^{-1}$, so these values are at the lower edge of the priors. In comparison the $K_{\rm zz}$ values used in our microphysical model range between $10^{7}...10^{9}\,{\rm cm^{2}\,s^{-1}}$, values comparable to the directly imaged brown dwarf result from 
 \cite{22MiBiPa}. Hence with a reduction by a factor of $10^{-2}$, we find some agreement between the retrieved $K_{\rm zz}$ and a microphysical model. Although, the deeper cloud deck (as shown in Section~\ref{subsec:Cloud_opt_depth}) still is higher than  shown in Figure~\ref{fig:Kzz_pressure}. We again re-iterate that the mechanism for such reduced mixing is uncertain.
 
Figure~\ref{fig:corner_cloud_formation} shows the free parameter, the integrated cloud particle nucleation rate $\dot\Sigma\,[{\rm g\,cm^{-2}\,s^{-1}}]$ is not particularly well-constrained by the retrieval, but a lower value is preferred. The nucleation rate, $J_{\rm i}$ [cm$^{-3}$s$^{-1}$] for i=TiO$_2$, SiO, KCl, NaCl,  in our kinetic cloud formation model is not a parameter, but rather we calculate the local nucleation rate, (see Section~\ref{s:ap}) according to the local thermodynamic properties ($T_{\rm gas}, p_{\rm gas}$). The integrated cloud particle nucleation rate can be calculated a

\begin{equation}
    \dot\Sigma = \int \sum_{i} m_{i}J_{i} {\rm d}z.
\end{equation}

Where the summation is over the four nucleation species considered ($i=$\ce{TiO2},\ce{SiO},\ce{NaCl},\ce{KCl}), with the mass of individual monomer species $m_{i}$ and the respective nucleation rates $J_{i}$. The integration depends on the computational volume of the GCM. The cloud extension is hence, a hidden within the $\dot\Sigma$ parameter. Using this, the terminator profiles (both morning and evening) from the GCM give integrated nucleation rates, of between $1.49\times 10^{-15}...4.56\times10^{-14}\,{\rm g\,cm^{-2}\,s^{-1}}$. This is in broad agreement with the retrieved values for the integrated nucleation rate, although the retrieval does favour slightly lower values 5~$\times$~$10^{-17}$ to 6~$\times$~$10^{-14}\,{\rm g\,cm^{-2}\,s^{-1}}$, within 1~$\sigma$ bounds. Examining Figure~\ref{fig:corner_cloud_formation}, we see that the lower retrieved values of $\dot\Sigma$ are towards the bottom of the prior region (lower limit of $10^{-17}\,{\rm g\,cm^{-2}\,s^{-1}}$). This is because the retrieval is essentially trying to minimise the formation of clouds in an atmosphere that, as the kinetic cloud formation model shows, should form clouds.
 
Another factor which could influence the observed spectra is the porosity of the cloud particles. The cloud particles in our retrievals are all assumed to be non-porous (porosity~=~0). Figure~\ref{fig:cloud_formation_Kzz} shows, however, that increasing porosity of cloud particles acts in a similar way to reducing the vertical mixing of cloud particles. The clouds become more transparent, reducing their optical thickness as porosity is increased. This particularly affects wavelengths shorter than around 2~$\mu$m. This trend agrees with the deeper cloud deck pressure found in Section~\ref{subsec:Cloud_opt_depth} when including the effects of cloud porosity. However the effect here is greatly reduced when compared to Figure~\ref{fig:Optical_depth} (lower left), due to the already deeper cloud in the best fit constrained model. The effect of cloud particle porosity has been explored further in works such as \cite{20SaHeMi}. The plausibility of highly porous cloud particles has been discussed in Section~\ref{subsec:Cloud_opt_depth}.

\begin{figure}
    \centering
      \includegraphics[width=\linewidth]{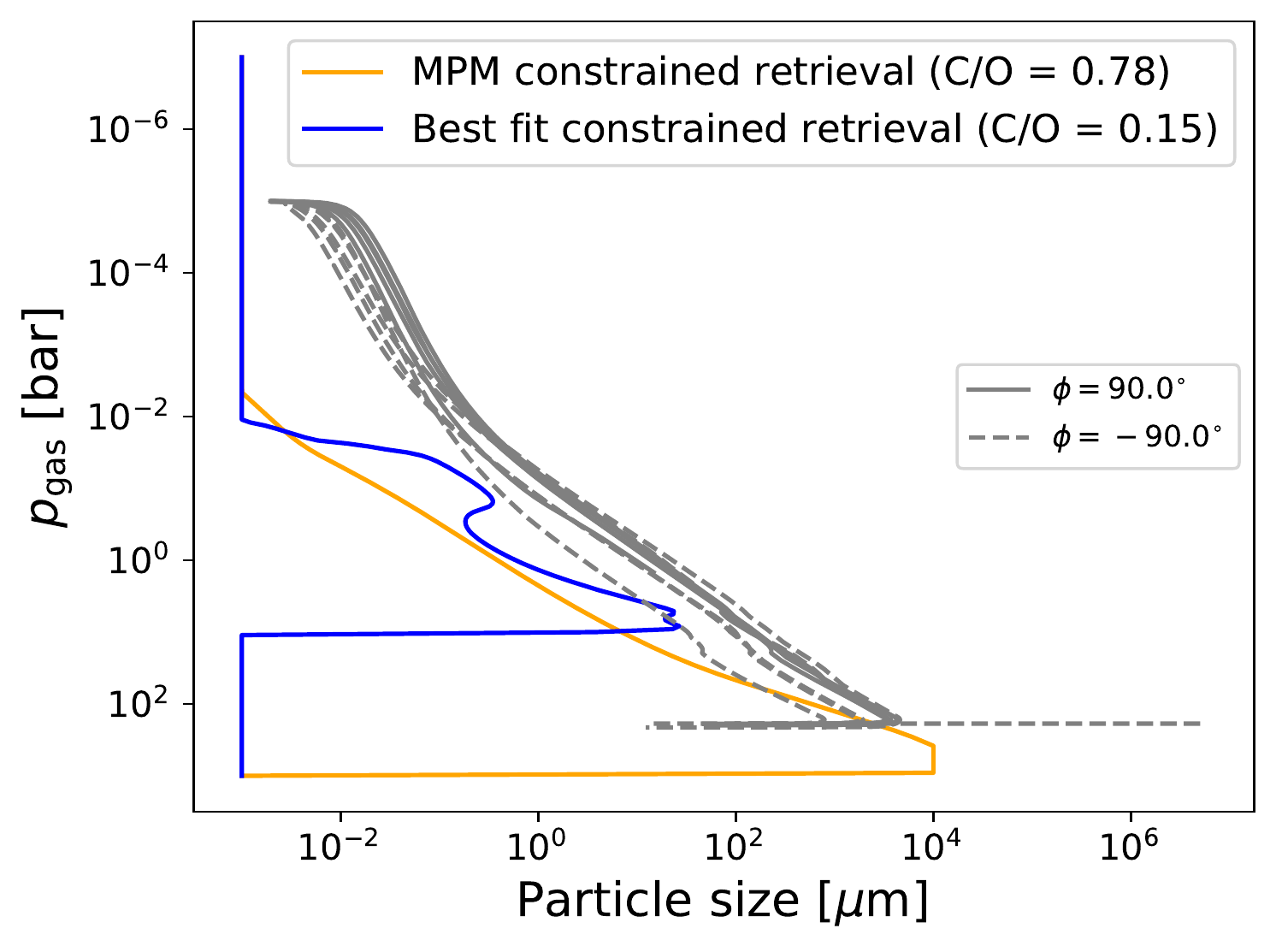}
    \caption{Average particle sizes from the constrained retrieval with ARCiS for the MPM (yellow) and the best fit (blue) models. Over-plotted are surface averaged particle sizes $\langle a \rangle_{\rm A}$ results from the kinetic cloud formation model (grey), for the evening ($\phi=90.0^{\circ}$, solid) and morning terminators ($\phi=-90.0^{\circ}$, dashed).}
    \label{fig:particle_sizes}
\end{figure}

Figure~\ref{fig:particle_sizes} shows the average particle size retrieved by ARCiS for the `best fit' and `MPM' solutions as well as the surface average particle size from the kinetic cloud formation model (Appendix A, Eq.~A.6 \citealt{2020A&A...641A.178H}). The average cloud particle size from the kinetic cloud formation are larger than the retrieved average particle sizes at all pressures in the atmosphere. The kinetic cloud formation model experiences condensational growth for the cloud particles right from the top of the GCM domain $10^{-5}\,{\rm bar}$, and hence larger particle sizes. This contrasts with the results from ARCiS, where for both solutions growth of average particle size does not start until deeper in the atmosphere. Both the kinetic cloud formation model and the ARCiS parameterised model use cloud condensation nuclei (CCN), the first step of cloud formation of size $\sim 10^{-3}\,\mu{\rm m}$. 
 
Figure~\ref{fig:comp_cloud_formation_no_cloud_ret} demonstrates that the retrieval result both (constrained and un-constrained) have large uncertainties in the retrieved pressure-temperature profiles. A temperature inversion could indicate clouds present deeper in the atmosphere, acting to heat the upper layers and thus making them too hot for further cloud formation. This offers another explanation for why the atmosphere appears relatively cloud free from observations, but could still contains an abundance of clouds as predicted by our kinetic models. 
 
\subsection{Discussion}
\label{subsec:discussion}

The compatibility of combining ground- and space-based data has been questioned in the case of WASP-96b by \cite{20YiChEd}, who find a significant offset between HST and VLT observations. The offset was also addressed by \cite{2022MNRAS.tmp.1485N}, who performed a combined analysis of various observations (VLT, HST, and Spitzer), and also determine an offset between the HST and VLT observations. This dataset (including their offset) is the one used in the retrievals of the present work.
 
\cite{20YiTsWa} further demonstrate the potential issues with combining data from instruments such as HST/WFC3 and Spitzer/IRAC which have no overlapping wavelength coverage. As the Spizter data photometry points come with large uncertainties, use of them in retrievals to quantify abundances of species such as CO or CO$_2$ can lead to biases in retrievals, particularly when constraining elemental abundances such as C/O. This has been highlighted previously by works such as \cite{19IrPaTa.wasp43b}. We therefore caution our retrieved value of C/O and await reduced spectra observed using JWST for further analysis. Our results which include cloud formation tentatively show two potential regions of parameter space which fit the observed pre-JWST data well: one with high C/O and a lower atmospheric temperature, and one with low C/O and a higher atmospheric temperature (in the region probed by the observations). The presence of large H$_2$O features in the region covered by the JWST observations could push towards the low C/O with higher atmospheric temperature solution. We note that the best-fit analysis and retrieval of \cite{2018Natur.557..526N} find a sub-solar C/O in both cases. These retrieved C/O are bulk C/O (i.e. the C/O available before clouds are allowed to form), and therefore a retrieved C/O without taking cloud formation into account will be higher than this bulk value. The bulk C/O is useful to be able to make links to planet formation; a parameterised planet formation model such as that of \cite{21KhMiDe} can be coupled with retrieval codes like ARCiS. 

In Figure~\ref{fig:cloud_abundances} we show the cloud species formed by the two solutions found in our retrieval in the two upper panels. On the left column we show the `best fit solution' with C/O~=~0.15 and a higher atmospheric temperature structure, and on the right the `MPM' solution with C/O~=~0.78 and a lower atmospheric temperature structure. The corresponding retrieved volume mixing ratios (VMR) of the gas-phase molecular species are given in the panels below.  In the `MPM' case, the high C/O (planetary bulk C/O) indicates not much oxygen available to form clouds. Condensates such as Fe, SiO$_2$, Al$_2$O$_3$ (corundum), TiO$_2$ and  VO are able to form, and the Na/K silicate and Mg/Ca silicate condensates do not have enough oxygen left to be able to form even though the temperatures would allow them to do so. The `best fit' case, on the other hand, has an abundance of oxygen in the atmosphere (low C/O). This allows the lower-temperature silicate clouds to form. The higher temperature profile does mitigate how much can form to some extent. Even with the Na/K silicate clouds that do form, it can be seen from the bottom left panel of Figure~\ref{fig:cloud_abundances} that there is still an abundance of gas-phase Na and K available to produce the strong atomic features seen in the observed spectra.

\section{Conclusion}

Based on our hierarchical modelling approach that combines a 3D GCM atmosphere solution for WASP-96b with a kinetic, non-equilibrium formation  model for mixed-material cloud particles, we suggest that WASP-96b is not cloud-free as previously instigated based on retrieval approaches. This suggestion is supported by applying the ARCiS retrieval framework to the same HST, Spitzer and VLT data which shows that cloudy solutions do reproduce the observed spectra. More than one cloudy solution provides good fits to the observational data such that retrieved solutions may not be unique, requiring more physical input which is an important step towards discussing retrieval results from present and future JWST data.

\smallskip\noindent
Clouds in WASP-96b would cause the following effects within the JWST wavelength range:
\begin{itemize}
\item The cloud top varies with wavelength within the JWST NIRSpec and NIRISS spectral ranges for at least 1 orders of magnitude in pressure.

\item Clouds become optically thick at different pressures dependent on wavelengths. To achieve optically thin clouds down to $p=10^{-2}\,{\rm bar}$, as implied by the sodium feature at $\sim0.6\, \mu{\rm m}$, a moderate mixing efficiency is required.

\item The long wavelength end of NIRSpec and short end of MIRI may probe atmospheric asymmetries between the limbs of the terminator on WASP-96b.

\item WASP-96b could only be cloud free in the unlikely case of a truly static atmosphere, and only cloud free in the region probed by observations in the case of extremely low mixing efficiency or if a temperature inversion confines the clouds to higher pressures.
\end{itemize}

\begin{acknowledgements}
Ch.H., M.M., L.C., and A.D.S. acknowledge funding from the European Union H2020-MSCA-ITN-2019 under Grant Agreement no. 860470 (CHAMELEON). K.L.C. acknowledges funding from STFC, under project number  ST/V000861/1. D.S. acknowledges financial support
from the Austrian Academy of Science. The authors thank David Lewis for their helpful discussions on the manuscript.
\end{acknowledgements}

\bibliographystyle{aa}
\bibliography{reference.bib}

\begin{appendix}
\onecolumn
\section{GCM parameters}
\newcommand{\GCMtable}[2]{ #1 & #2\\}
\begin{table}[!ht]
    \centering
    \caption{Model parameters for the GCM used to produce 1D profiles for WASP-96b}
    \begin{tabular}{cc}
        \hline
        \hline
         \GCMtable{Parameter}{Value}
         \hline
         \GCMtable{Dynamical time-step $\Delta t$}{$25\,{\rm s}$}
         \GCMtable{Radiative time-step $\Delta t_{\mathrm{rad}}$}{$100\,{\rm s}$}
         \GCMtable{Stellar Temperature$^{1}$ ($T_{*}$)}{$5540\,{\rm K}$}
         \GCMtable{Stellar Radius$^{1}$ ($R_{*}$)}{$1.05\,R_{\rm sun}$}
         \GCMtable{Semi-major axis$^{1}$ ($a_{\rm p}$)}{$0.0453\,{\rm au}$}
         \GCMtable{Substellar irradiation temperature$^{2}$ ($T_{\rm irr}$)}{$1819\,{\rm K}$}
         \GCMtable{Planetary Radius$^{1}$ ($R_{\rm p}$)}{$1.2\,R_{\rm Jup}$}
         \GCMtable{Specific heat capacity at constant pressure$^{3}$ ($c_{\rm p}$)}{$13784\,{\rm J\,kg^{-1}\,K^{-1}}$}
         \GCMtable{Specific gas constant$^{3}$ ($R$)}{$3707\,{\rm J\,kg^{-1}\,K^{-1}}$}
         \GCMtable{Rotation period ($P_{\rm rot}$)}{$3.4\,{\rm days}$}
         \GCMtable{Surface gravity ($g$)}{$826\,{\rm cm\,s^{-2}}$}
         \GCMtable{Lowest pressure ($p_{\rm top}$)}{$10^{-5}\,{\rm bar}$}
         \GCMtable{Highest pressure ($p_{\rm bottom}$)}{$700\,{\rm bar}$}
         \GCMtable{Vertical Resolution ($N_{\rm layers}$)}{$47$}
         \GCMtable{Wavelength Resolution$^{4}$ (S1)}{$11$}
        \hline
        \hline
    \end{tabular}
    \\ Notes: 1. Values taken from \cite{14HeAnCa}, 2. Calculated using Eq.~1 from \citep{10Guillot.exo}, 3. Inferred using petitRADTRANS equilibrium package, 4. Same as \cite{Kataria2013}, benchmarked in Appendix B. of \cite{2022arXiv220209183S}
    \label{tab:GCM_params}
\end{table}

\newpage
\section{Posterior Distributions}

\begin{figure}[ht]
    \centering
    \includegraphics[width=\linewidth]{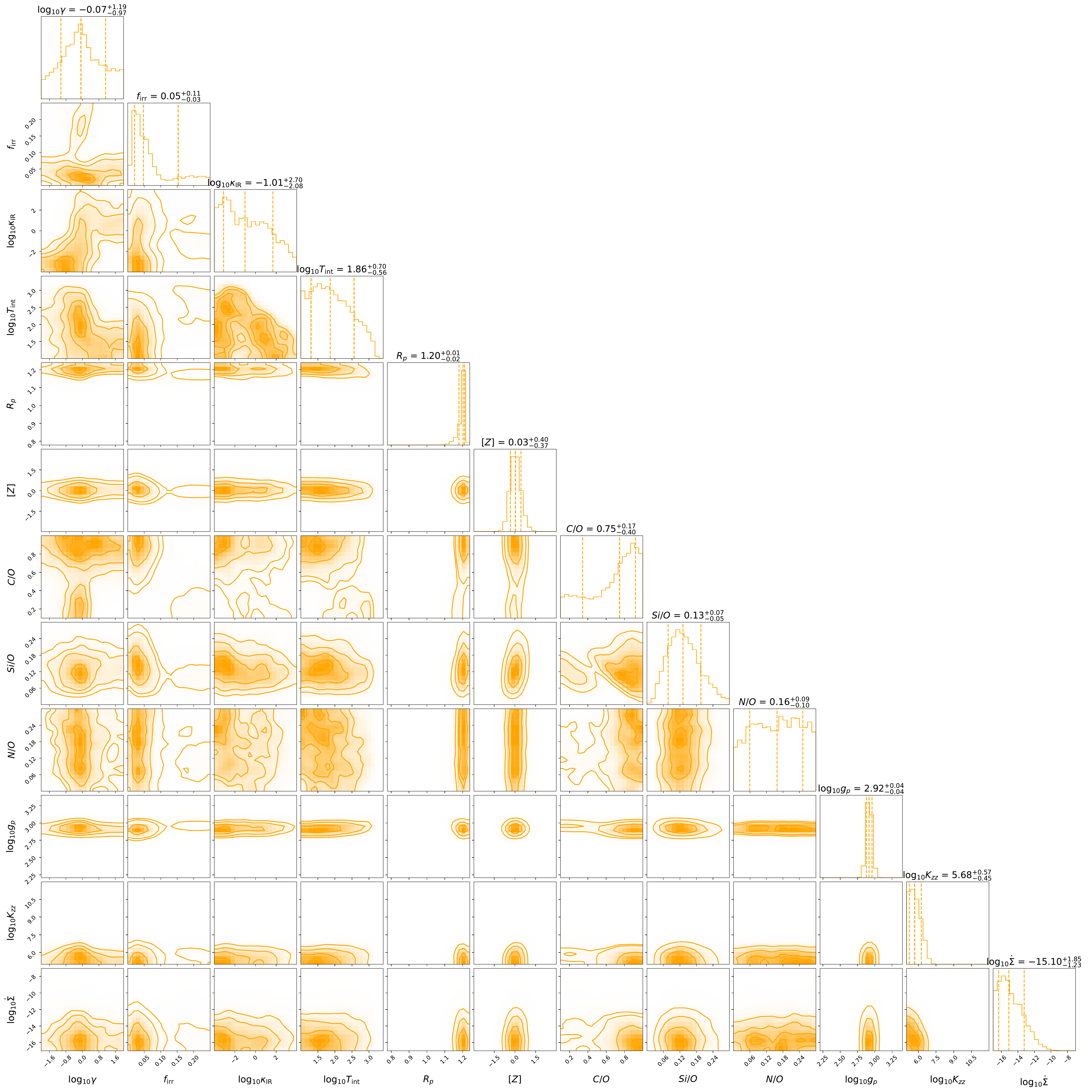}
    \caption{The posterior distributions for the equilibrium chemistry retrieval of \cite{2022MNRAS.tmp.1485N} pre-JWST observations with cloud formation included (as described in \cite{18OrMi.arcis}). See Fig.~\ref{fig:comp_cloud_formation_no_cloud_ret} (right)
     for the PT-profile, which was computed using retrieved parameters for a PT-profile based on that of \cite{10Guillot.exo}.
     All parameters are described in Table~\ref{t:ret_pars}.}
    \label{fig:corner_cloud_formation}
\end{figure}

\begin{figure}[ht]
    \centering
     \includegraphics[width=\linewidth]{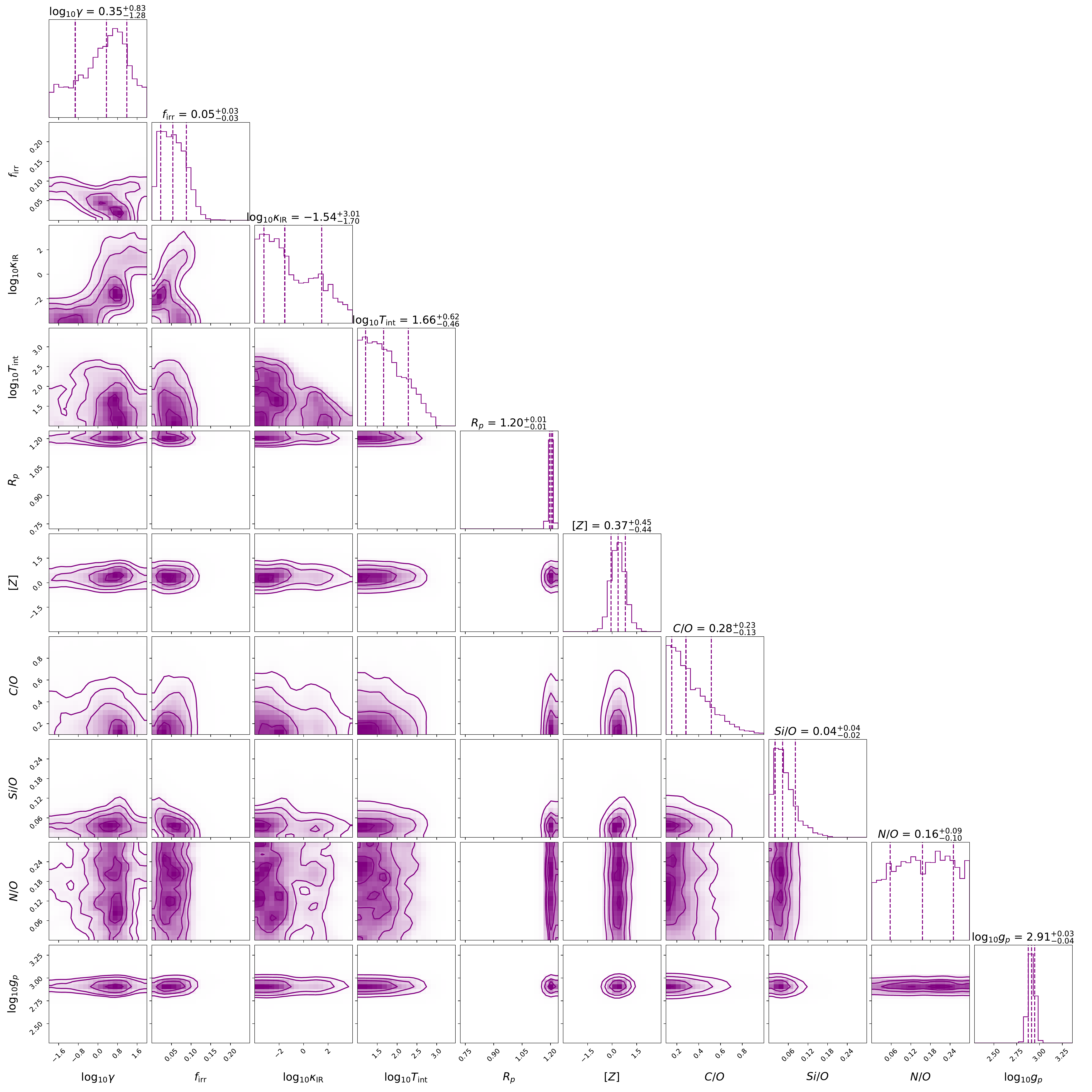}
    \caption{The posterior distributions for the equilibrium chemistry retrieval of \cite{2022MNRAS.tmp.1485N} pre-JWST observations with no cloud parameters included.}
    \label{fig:corner_no_clouds}
\end{figure}

\newpage
\onecolumn
\section{Additional retrieval figures}
\begin{figure}[ht]
    \centering
     \includegraphics[width=0.45\linewidth]{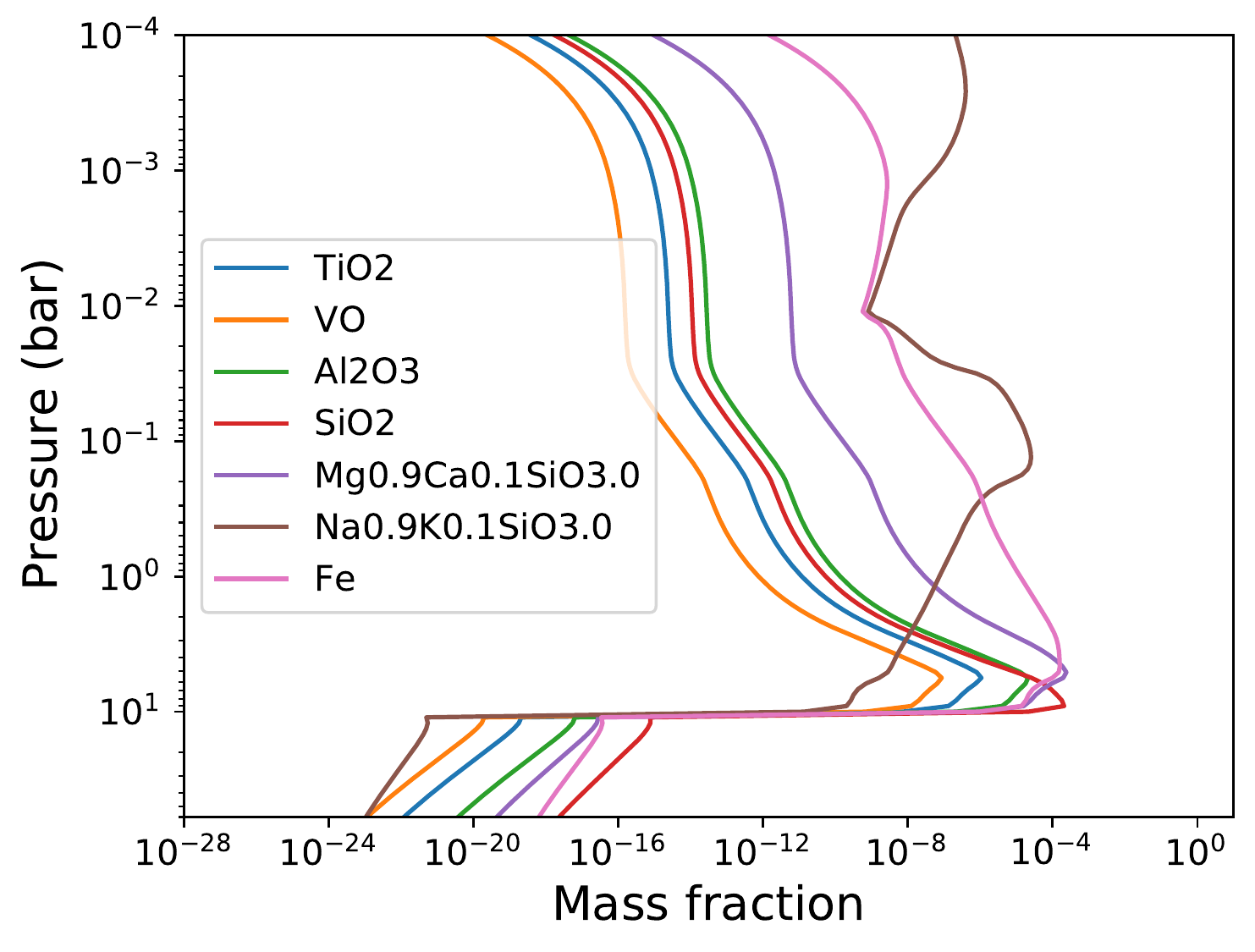}
      \includegraphics[width=0.45\linewidth]{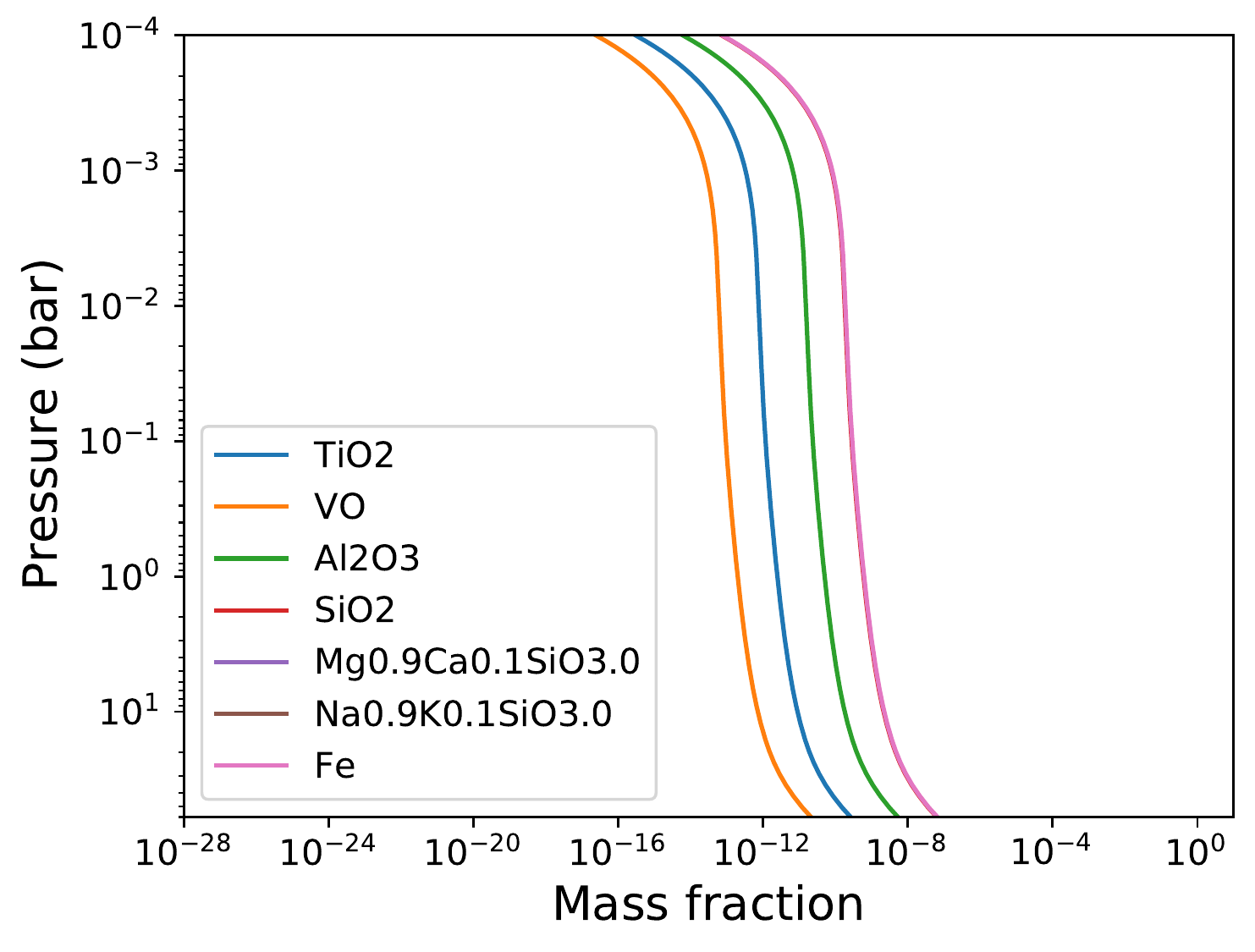}
      \includegraphics[width=0.45\linewidth]{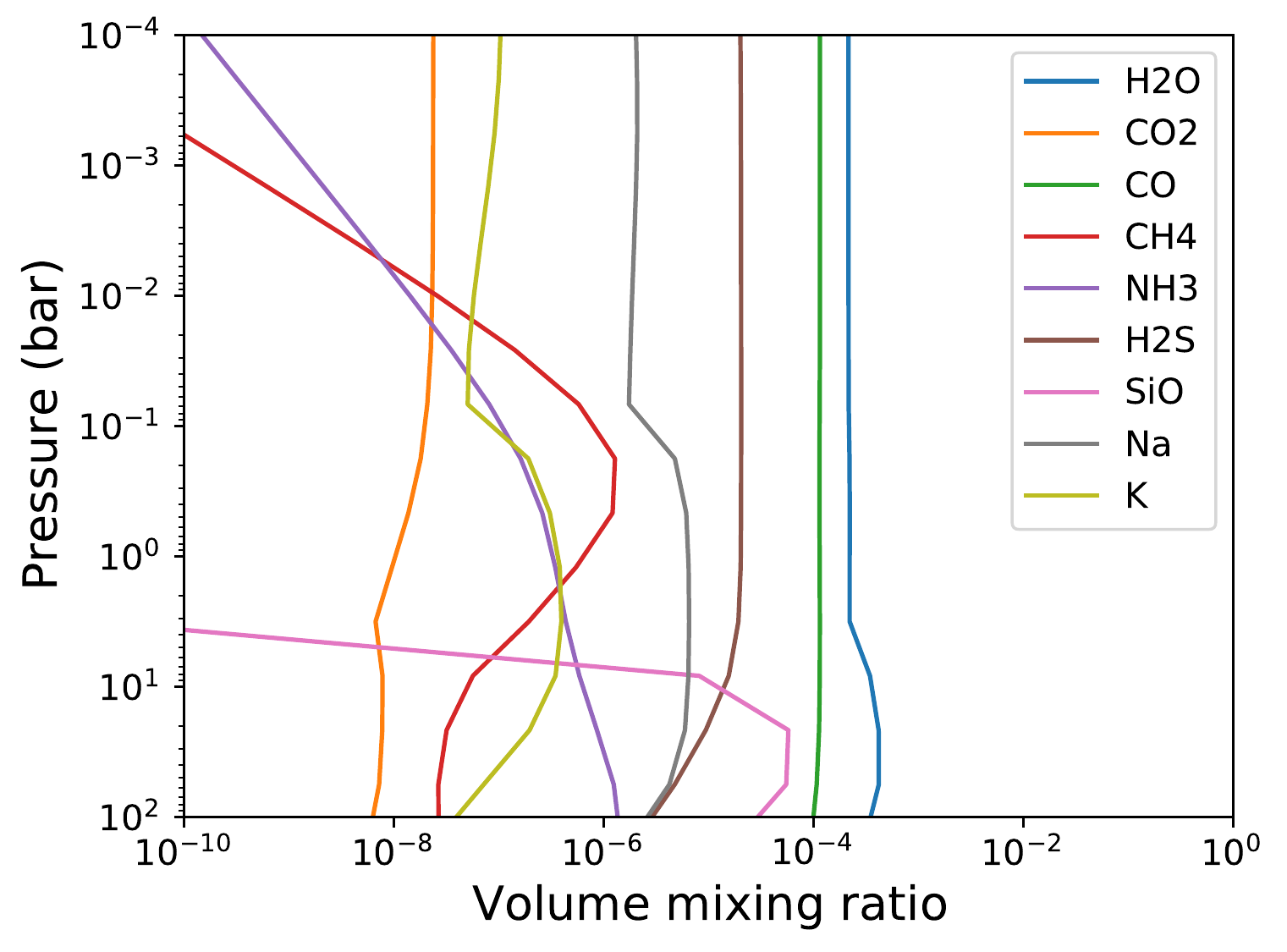}
      \includegraphics[width=0.45\linewidth]{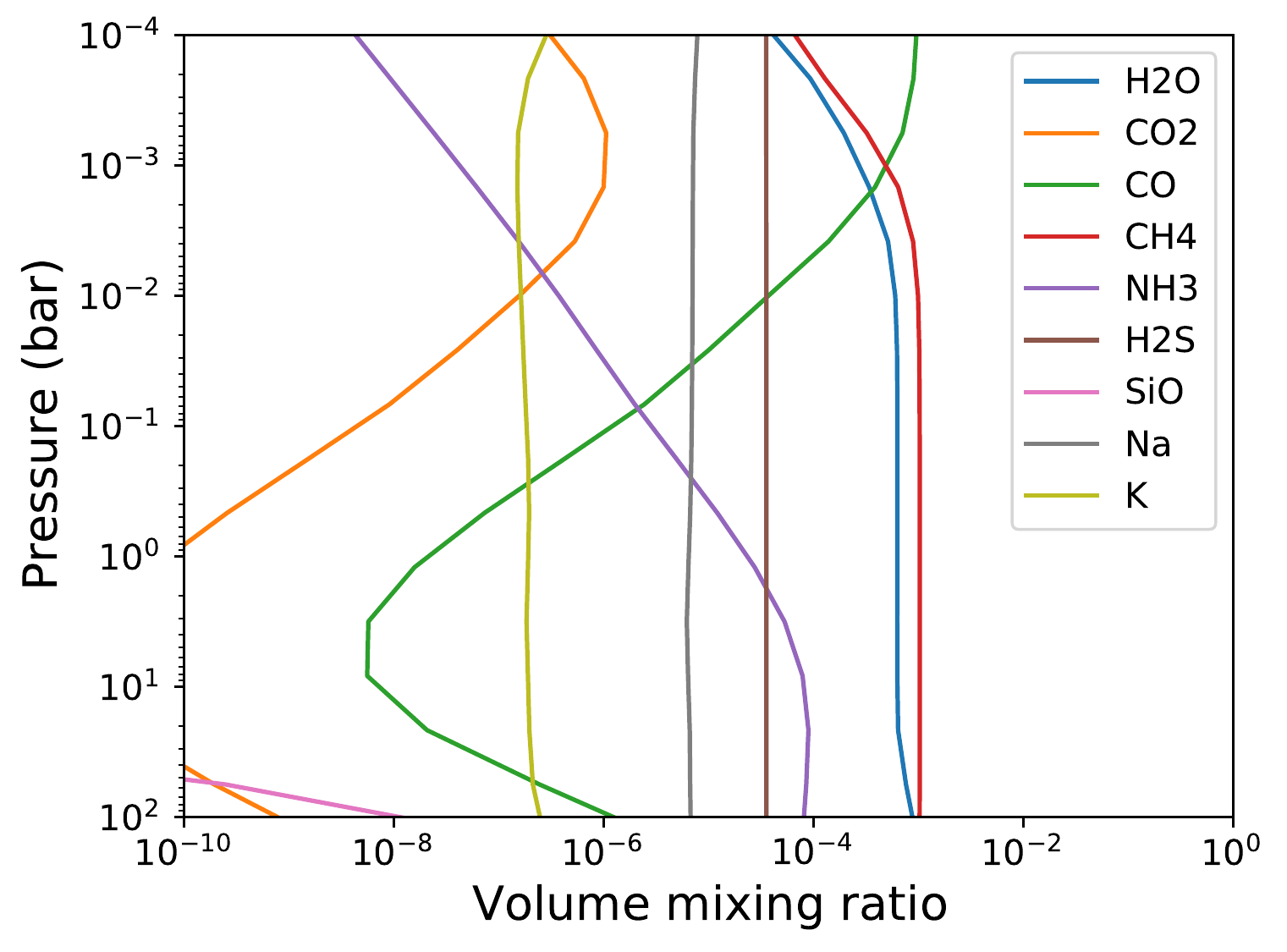}
    \caption{\textbf{Top:} Mass fractions (solid mass / gas mass) for the main cloud species formed in the constrained cloud formation retrievals. \textbf{Left:} the `best fit' solution (C/O~=~0.15, higher temperature). \textbf{Right:} the `MPM' solution (C/O~=~0.78, lower temperature). Cloud species included in the formation models but not formed in high-enough amounts to be shown on these plots: H$_2$O ice, amorphous carbon, SiC. Underneath are the corresponding volume mixing ratios of the most abundant molecular species included in the retrievals.}
    \label{fig:cloud_abundances}
\end{figure}

\begin{figure}[ht]
    \centering
    \includegraphics[width=0.45\linewidth]{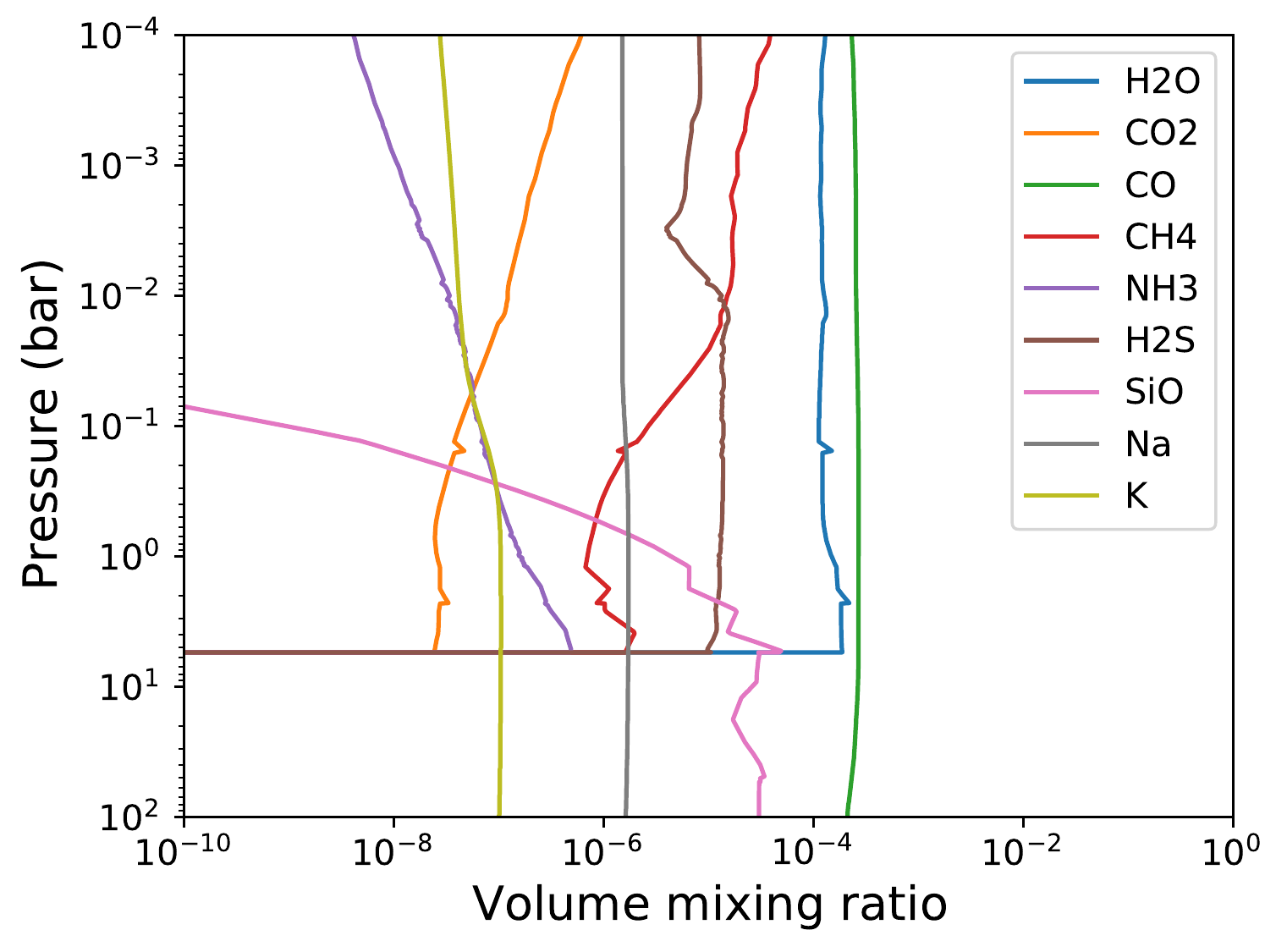}
    \includegraphics[width=0.45\linewidth]{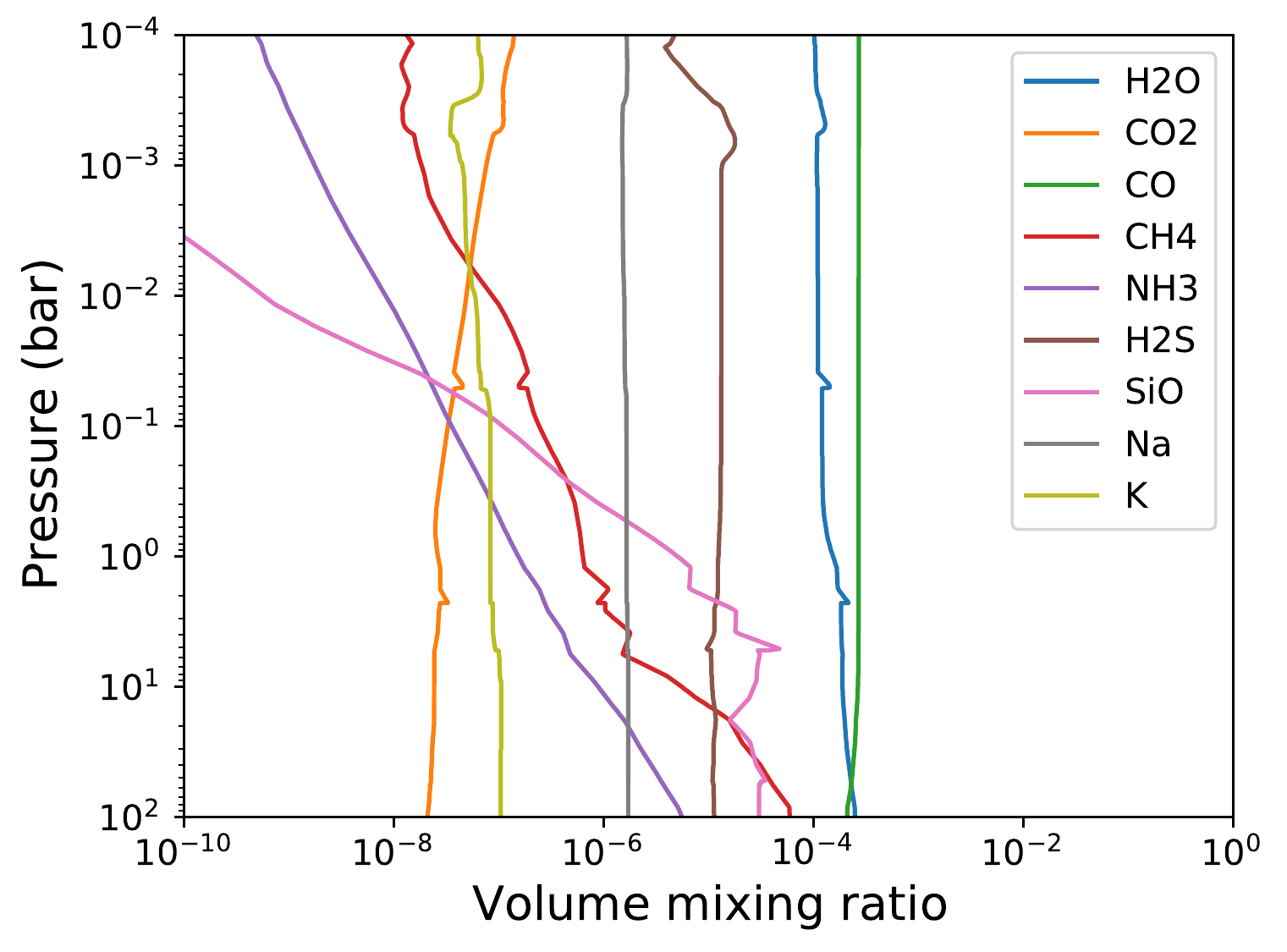}
    \caption{Mixing ratios of a selection of the most abundant molecules at the equator of the morning (left) and evening (right) terminator predicted by the GCM and cloud formation models of this work.}
    \label{fig:mixrat_GCM}
\end{figure}

\end{appendix}

\end{document}